\documentclass[letterpaper,english,5p,sort&compress]{elsarticle}
\usepackage[T1]{fontenc}
\usepackage[latin9]{inputenc}
\usepackage{color}
\usepackage{amsmath}
\usepackage{amssymb}
\usepackage{graphicx}

\makeatletter

\newcommand{\lyxdot}{.}

\newenvironment{lyxlist}[1]
	{\begin{list}{}
		{\settowidth{\labelwidth}{#1}
		 \setlength{\leftmargin}{\labelwidth}
		 \addtolength{\leftmargin}{\labelsep}
		 }}
	{\end{list}}

\usepackage{amsmath}
\DeclareMathOperator*{\argmax}{argmax}
\DeclareMathOperator*{\argmin}{argmin}

\@ifundefined{showcaptionsetup}{}{%
 \PassOptionsToPackage{caption=false}{subfig}}
\usepackage{subfig}
\makeatother

\usepackage{babel}
\begin{document}
\global\long\def\hc{\text{h.c.}}%
 
\global\long\def\re{\text{Re}}%
 
\global\long\def\im{\text{Im}}%
 
\global\long\def\K{\mbox{ K}}%
 
\global\long\def\meV{\mbox{ meV}}%
 
\global\long\def\eV{\mbox{ eV}}%
 
\global\long\def\hf{\!\,_{2}F_{1}}%
 
\global\long\def\pr{\prime}%
 
\global\long\def\prpr{\prime\prime}%
 
\global\long\def\sgn{\text{sgn}}%
\global\long\def\amin{\argmin}%
\global\long\def\amax{\argmax}%

\begin{frontmatter}{}

\title{Microstructural and compositional design principles for Mo-V-Nb-Ti-Zr
multi-principal element alloys: a high-throughput first-principles
study}

\author[a]{Zhidong Leong}

\ead{leong\_zhidong@ihpc.a-star.edu.sg}

\author[b,c]{Upadrasta Ramamurty}

\ead{uram@ntu.edu.sg}

\author[a]{Teck Leong Tan\corref{cor}}

\cortext[cor]{Corresponding author}

\ead{tantl@ihpc.a-star.edu.sg}

\address[a]{Institute of High Performance Computing, Agency for Science, Technology
and Research, Singapore 138632, Singapore}

\address[b]{School of Mechanical and Aerospace Engineering, Nanyang Technological
University, Singapore 639798, Singapore}

\address[c]{Institute of Materials Research and Engineering, Agency for Science,
Technology and Research, Singapore 138634, Singapore}
\begin{abstract}
Due to the vast compositional space of multi-principal element alloys
(MPEAs), the rational design of MPEAs for optimized microstructures
is difficult. Therefore, a high-throughput first-principles study
of Mo-V-Nb-Ti-Zr, a refractory MPEA, was conducted to gain insights
into the underlying microstructures. Using Monte-Carlo simulations
powered by cluster expansion, we uncover the principles governing
the MPEA's microstructures across a large compositional space that
includes non-equiatomic compositions and encompasses the constituent
binaries, ternaries, and quaternaries. In the spirit of Hume-Rothery
rules for complete solid solubility, we present a quantitative expression
for predicting solid solution formation from the composition. Within
a consistent framework, our results reproduce the microstructural
observations (solid solution vs. segregation) from numerous experiments
and provide microstructural predictions for unexplored regions in
the compositional space. Our work illuminates the MPEA's microstructures
in terms of the separation and clustering tendencies of the elements,
presenting a set of simple but powerful design principles for future
experiments to rationally design MPEAs with the desired microstructures
for superior mechanical properties.
\end{abstract}

\end{frontmatter}{}

\section{Introduction}

An extension of the original design concept of high-entropy alloys
(HEAs), multi-principal element alloys \linebreak{}
(MPEAs) are alloys with three or more principal components of comparable,
but not necessarily equal, compositions \citep{Miracle2019}. In pursuing
rational design strategies for MPEAs, the field of MPEAs focuses on
stabilizing the solid-solution phase via maximizing the configurational
entropy, as well as using secondary phases to optimize the strength-ductility
combinations. However, the exceptional tunability of MPEAs is a double-edged
sword: while it can lead to physical properties superior to those
of conventional alloys \citep{George2019,Miracle2017,Zhao2017}, it
complicates the understanding of MPEAs and hence hinders rational
design. 

For example, refractory MPEAs, such as Mo-V-Nb-Ti-Zr, have been studied
for their high-temperature strength \citep{Senkov2020,Senkov2018a,Senkov2010,Juan2015}
and biomechanical compatibility \citep{Chui2020,Ching2020,Yuan2019,Todai2017,Nagase2020}
for aerospace and biomedical applications, respectively. However,
insights into the physics of refractory MPEAs remain limited \citep{Senkov2018a},
because experiments are often restricted to just a small number of
compositions within the huge compositional space, which has vastly
expanded in MPEA design. With additive manufacturing, there is also
rising interests in compositionally graded MPEAs \citep{Dobbelstein2019,Gwalani2019}
and compositionally inhomogeneous MPEAs \citep{Chang2019a}. Therefore,
a general high-throughput computational methodology that covers the
full compositional design space of MPEAs, including both equiatomic
and non-equiatomic compositions, would be timely. Such a study can
reveal insights into the thermodynamic origins of the MPEA microstructures,
guiding future experiments in the rational design of MPEAs with the
desired properties and microstructures.

In the study of MPEAs, information about the microstructures in relation
to temperature and composition is important for design purposes because
they impact mechanical properties \citep{Feng2017,Ding2018,Li2020,Chen2014,Varvenne2017,Kumar2019,Zhao2020}.
In particular, understanding the formation of solid solutions vis-\`{a}-vis
the precipitation of secondary phases is critical \citep{Senkov2018a,Basu2020}.
While Hume-Rothery rules \citep{Hume-Rothery1935,Hume-Rothery1952,Hume-Rothery1969}
are popular empirical rules for predicting solid solution formation
in conventional alloys, applying these rules to MPEAs is not always
reliable. Hence, the prediction of solid solution formation in MPEAs
remains a key priority of active research \citep{Senkov2015,Li2020a,Pei2020,Zeng2021}. 

Experimentally, X-ray diffraction (XRD) and various microscopy techniques
are often used to resolve the microstructures of MPEAs \citep{Senkov2018,Pradeep2013}.
The slow diffusion kinetics in MPEAs \citep{Tsai2014} does not allow
for the design of alloys with long-term microstructural stability.
In this context, the thermodynamic transition temperature, $T_{c}$,
is a key material characteristic. Above $T_{c}$, the solid-solution
phase is thermodynamically stable. Below $T_{c}$, the formation of
secondary phases becomes energetically favorable \citep{Zarkevich2007}.
Therefore, a compositional map of $T_{c}$, as well as the microstructural
information on phase formation, is highly desirable.

In this work, we elucidate the compositional dependence of the microstructures
of Mo-V-Nb-Ti-Zr MPEAs via a high-throughput first-principles study.
Using cluster expansion (CE) within Monte Carlo (MC) simulations,
we predict that, over the whole compositional design space, the single-phase
solid solution at high temperatures cools into two segregated secondary
phases: one Zr-rich and the other enriched with Mo and V, while Nb
and Ti remain randomly dispersed. Tracing the short-range ordering
within the microstructures reveals that this phase formation is predominantly
driven by Zr-V segregation tendency, followed by Mo-V clustering tendency.
Consequently, a higher V content accelerates secondary phase formation
and increases the $T_{c}$. In the spirit of Hume-Rothery rules, we
present a quantitative expression for predicting the $T_{c}$ and
identify compositions that preclude solid solution formation. The
accuracy of our methodology is validated by the agreement with the
experimental data available at various compositions and conditions
\citep{Zhang2012,Wu2015,Senkov2018}. Our work serves to guide the
rational design of MPEAs with the desired microstructures and mechanical
properties.

\section{Methods}

\subsection{Cluster expansion}

Cluster expansion (CE) \citep{Sanchez1984} is a popular computational
technique for modeling crystalline alloy properties. Based on the
generalized Ising model, the formation energies of alloy structures
are expanded in CE in terms of the effective cluster interactions
(ECIs) of the relevant atomic clusters (see \ref{sec:Appendix-CE}
for details). By fitting to energies calculated from first-principles,
ECIs between various elemental species in the alloy can be determined.
The result is a surrogate model that rapidly and accurately computes
the formation energy of any alloy structure, enabling high-throughput
thermodynamic studies of MPEAs.

While there exists abundant literature on using CE to study binary
and ternary alloys \citep{Connolly1983,Lu1991,Wolverton1992,Garbulsky1994,Ozolins1998,Muller2001,Zunger2002,vandeWalle2002c,Blum2004,Hart2005,Zarkevich2008,vandeWalle2009,Tan2012,Ravi2012,Nelson2013,Wrobel2015,VanderVen2018},
studies on quaternary \citep{Maisel2016,Feng2017,Fernandez-Caballero2019,El-Atwani2019}
and quinary \citep{Nguyen2017,Fernandez-Caballero2017} alloys remain
limited and are restricted to a small number of compositions. Practical
applications thus far include designing AlNiCo permanent magnets \citep{Nguyen2017}
and radiation resistant materials \citep{El-Atwani2019} using MPEAs. 

Here, we train our CE model of Mo-V-Nb-Ti-Zr (henceforth abbreviated
as MVNTZ) using the formation energies of 10265 alloy structures based
on density functional theory (DFT) (see \ref{sec:Appendix-DFT}).
Our CE is based on the bcc lattice because the solid solution phases
of all the ten constituent binaries are known to have such a lattice
structure. To ensure that our training data is representative, we
include binary to quinary structures covering a wide range of compositions.

\begin{figure}[t]
\subfloat[\label{fig:glasso-ECIs}]{\includegraphics[scale=0.4]{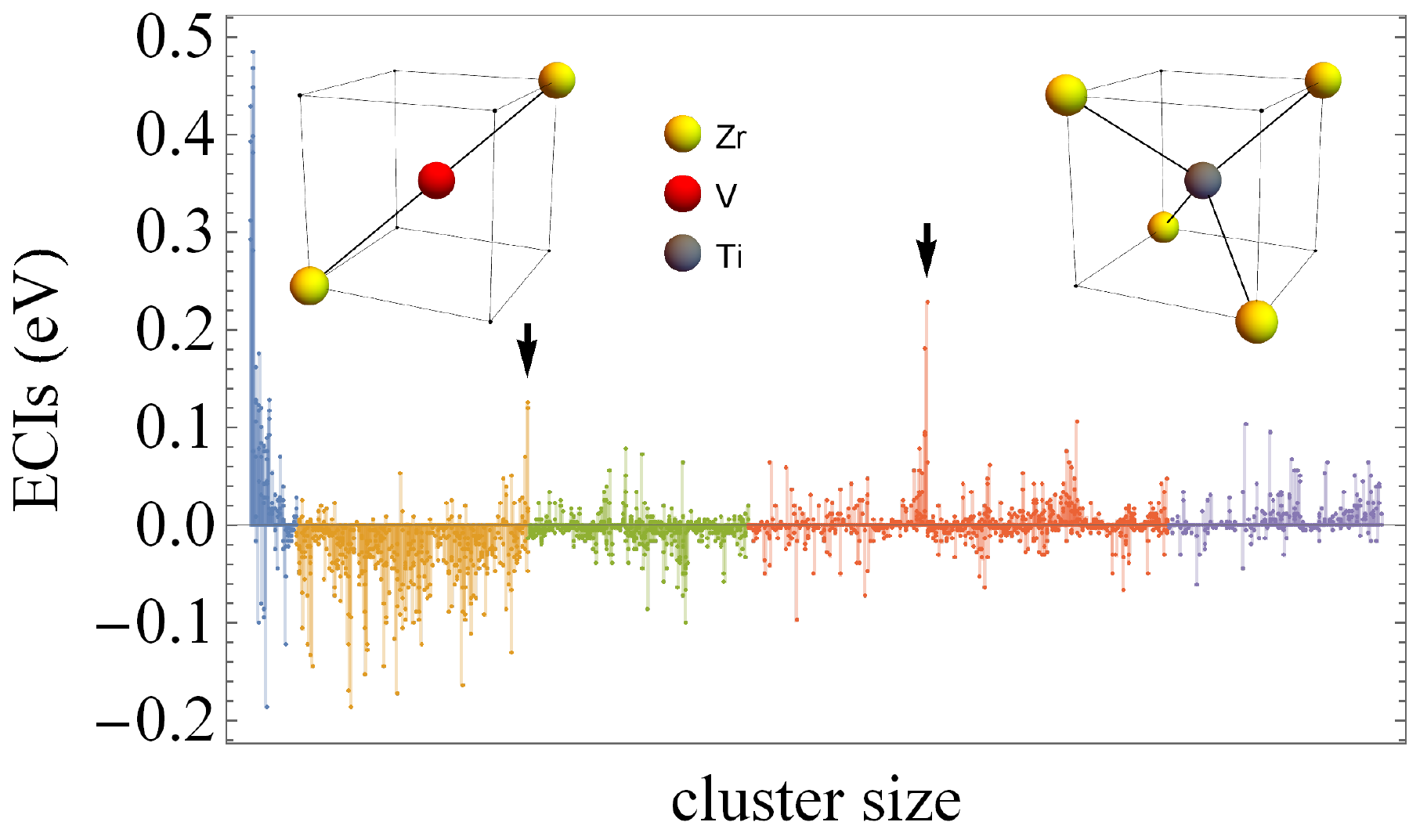}

}

\subfloat[\label{fig:CE-vs-DFT-energies}]{\includegraphics[scale=0.4]{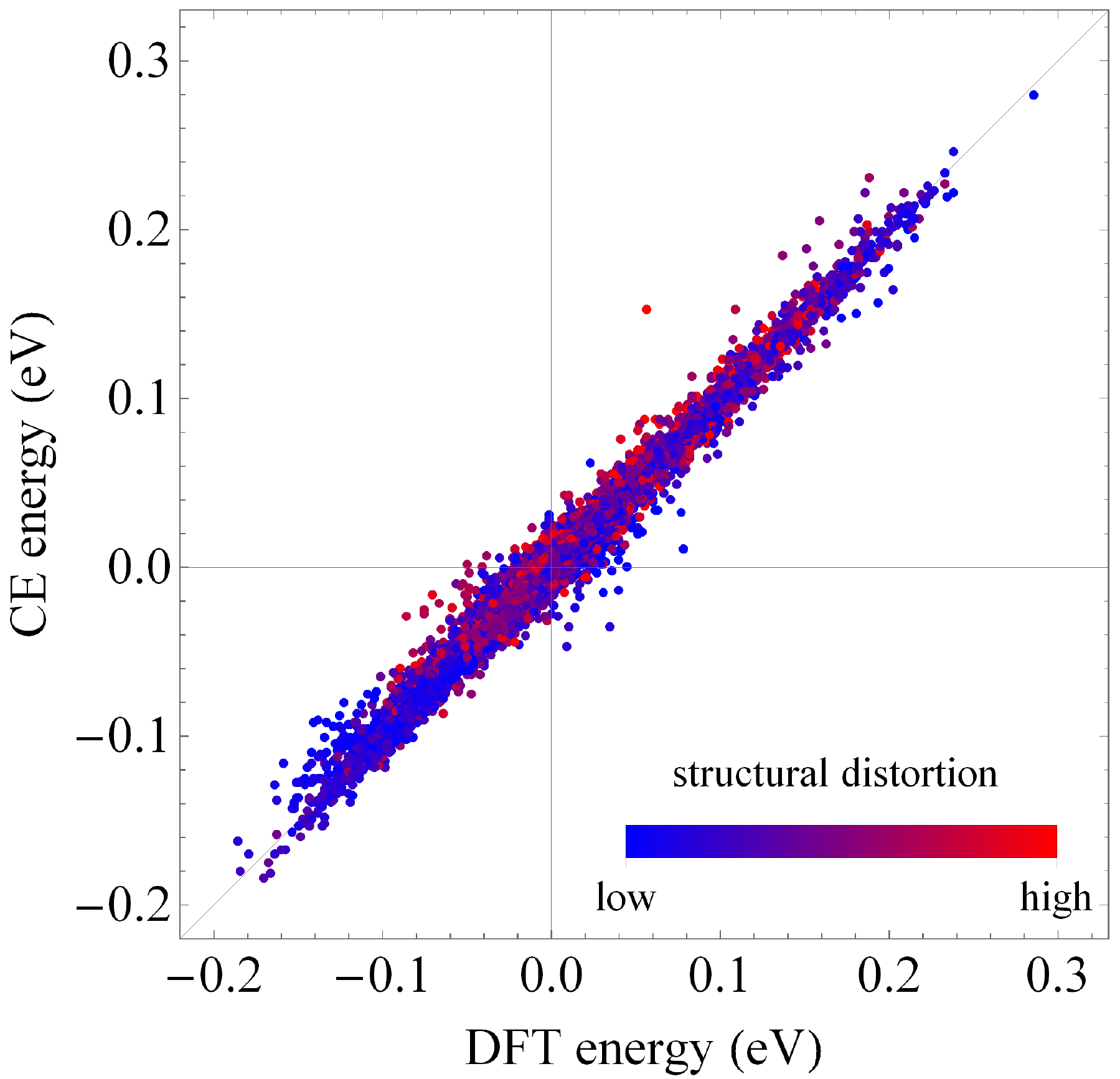}

}

\caption{Cluster expansion (CE) of MVNTZ, trained with the first-principles
formation energies of 10265 alloy structures. The cross-validation
score is $9.2\protect\meV$. (a) The fitted values of the 2911 ECIs.
Pairs, triplets, quadruplets, five-bodies, and six-bodies are colored
blue, orange, green, red, and purple, respectively. Insets: the three-
and five-body Zr-rich clusters corresponding to the large ECIs marked
by the arrows. (b) A plot of the predicted CE energies against the
DFT energies for the training structures. Each point is colored based
on the degree of structural distortion. }
\end{figure}
A key challenge in CE involves selecting the most predictive set of
atomic clusters as descriptors---an exhaustive search is computationally
prohibitive due to the large number of possible combinations, particularly
for MPEAs. Alloy structures with significant structural distortions
(e.g. due to large atomic size mismatch \citep{Ravi2012} or high
Zr content \citep{Tong2020}) are also known to complicate cluster
selection in CE. In a recent work \citep{Leong2019}, we developed
an efficient physics-based learning methodology for cluster selection.
Based on group lasso \citep{Yuan2006}, this methodology selects a
parsimonious set of atomic clusters in accordance with the physical
insight that if a cluster is selected, its subclusters should be too
\citep{Zarkevich2004}. Equivalently, a cluster's ECI is nonzero only
if the ECIs of all its subclusters are also nonzero. These selection
rules avoid spuriously large fitting parameters by redistributing
them among lower-order terms, resulting in more physical, accurate,
and robust CEs. A parsimonious CE consisting of just the essential
ECIs is important for accelerating the sampling of the MPEA's configurational
space during subsequent MC simulations. In this present paper, we
apply group lasso to study MPEAs.

Fig. \ref{fig:glasso-ECIs} shows the fitted values of the ECIs. From
an initial pool of 2911 atomic clusters consisting of two- to six-body
clusters, group lasso has selected 75\% of them for the CE; the ECIs
of the remaining 25\% have been set to zero to yield a parsimonious
model. The fitted ECIs exhibit good convergence---the magnitude of
the ECIs generally decreases with increasing cluster size. The spikes
among some of the triplets and larger clusters are necessary for capturing
the effects of structural distortion, particularly in Zr-rich structures
(see Fig. \ref{fig:distortion-correlation}). As the insets highlight,
these clusters tend to have high Zr content. Indeed, Fig. \ref{fig:unrelaxed-plots}
shows that these spikes vanish if we were to train our CE model with
the undistorted structures instead, i.e., structures where only the
unit cell parameters are relaxed (no local relaxation of atomic positions).
Accurately modeling the effects of structural distortions in MVNTZ
using higher-order interactions is important for the quantitative
agreement between CE predictions and DFT energies. Shown in Fig. \ref{fig:CE-vs-DFT-energies}
and further detailed in Fig. \ref{fig:CE-error-vs-distortion}, this
agreement is quantified by a cross-validation score of $9.2\meV$
and is good even for structures with high degrees of distortion. This
accurate surrogate model of MVNTZ is the key machinery that enables
the high-throughput MC simulations in this work, for producing thermodynamics
and microstructural information across compositions.

\subsection{Monte Carlo (MC) simulations}

We perform MC simulations at 505 compositions, where the concentration
of each elemental species has values $0,0.1,$\linebreak{}
$0.2,0.3,0.4,0.5$ with a nonzero amount of Zr. These compositions
span a space large enough to include almost all reasonable definitions
of MPEAs. Our canonical MC simulations are based on the Metropolis
algorithm as implemented in our Thermodynamic Toolkit (TTK) \citep{Zarkevich2007,Zarkevich2008}.
Each simulation contains $24\times24\times24$ atoms within the periodic
simulation box, using 500 equilibrium steps and 2000 sampling steps.
\textcolor{black}{These values are fixed and are deemed sufficient for
convergence by benchmarking our results for the constituent binary
systems against known phase diagrams.} At each composition, the simulation
begins in the solid solution phase at a sufficiently high temperature
and cools with a temperature step of $\Delta T=10\meV\sim116\K$,
for a total of 37 temperatures. 

\subsection{Short-range order (SRO) parameters}

The microstructures observed in MC simulations can be characterized
by using the Warren-Cowley short-range order (SRO) parameters \citep{Fontaine1971}
to track the elemental distributions. In our work, we focus on the
nearest-neighbor (NN) SRO parameter. For elemental species $i\neq j$,
the NN SRO parameter is given by 

\begin{equation}
\alpha_{ij}=1-\frac{p_{ij}}{c_{i}c_{j}},
\end{equation}
where $c_{i},c_{j}$ are the concentrations of species $i,j$, and
$p_{ij}$ is the probability of a NN atomic pair having species $i,j$
in the first and second sites, respectively. It follows that $\alpha=0$
indicates a fully random arrangement, $0<\alpha\leq1$ a segregation
tendency, and $\alpha<0$ a tendency for the species to cluster. 

\section{Results}

\subsection{Clustering/segregation tendency}

\begin{figure*}[t]
\begin{centering}
\subfloat[\label{fig:equicomp-SRO-vs-T}]{\begin{centering}
\includegraphics[scale=0.4]{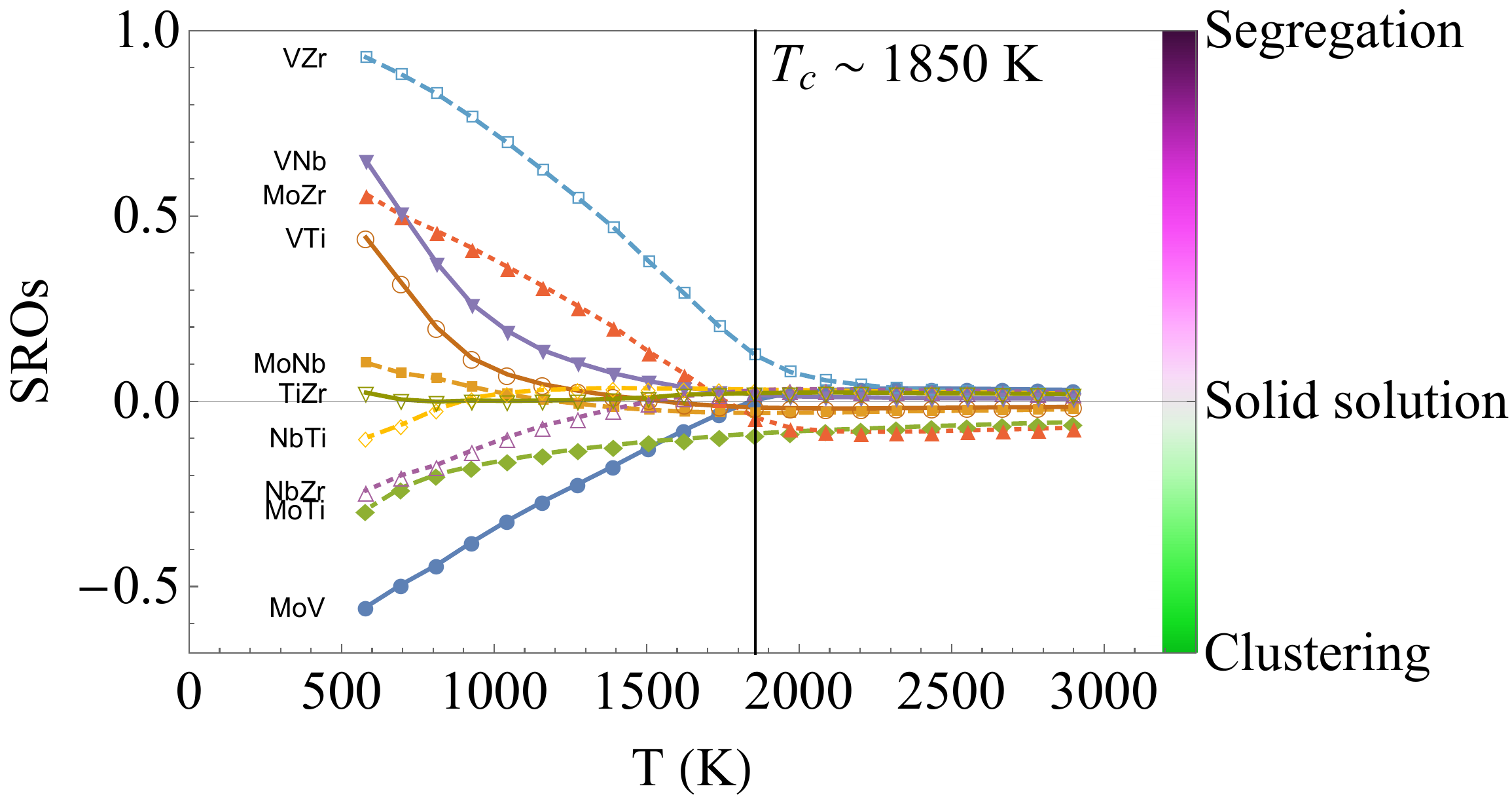}
\par\end{centering}
}
\par\end{centering}
\hfill{}\subfloat[$T\sim600\protect\K$ \label{fig:equicomp-low-T}]{\includegraphics[scale=0.35]{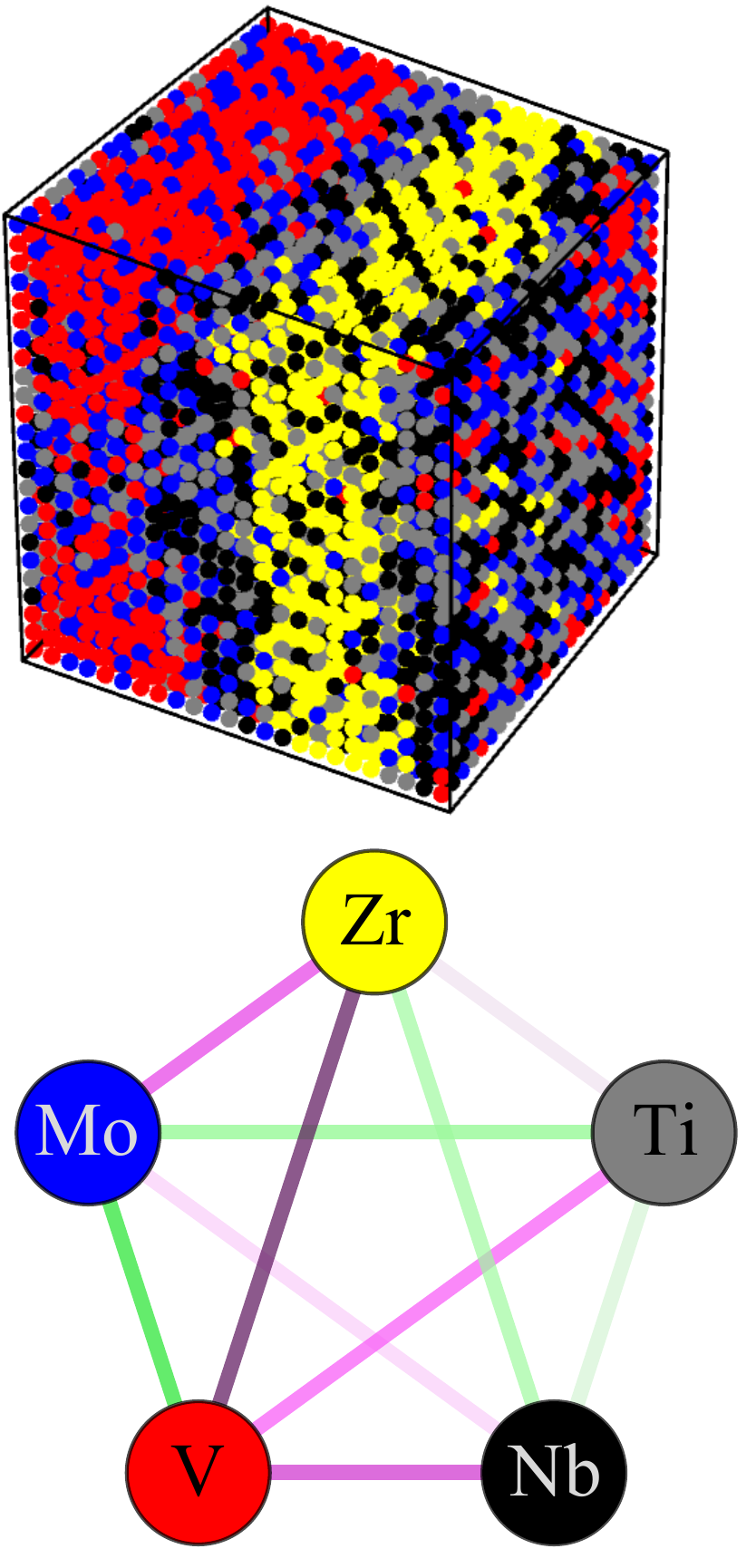}

}\hfill{}\subfloat[$T\sim1400\protect\K$ \label{fig:equicomp-mid-T}]{\includegraphics[scale=0.35]{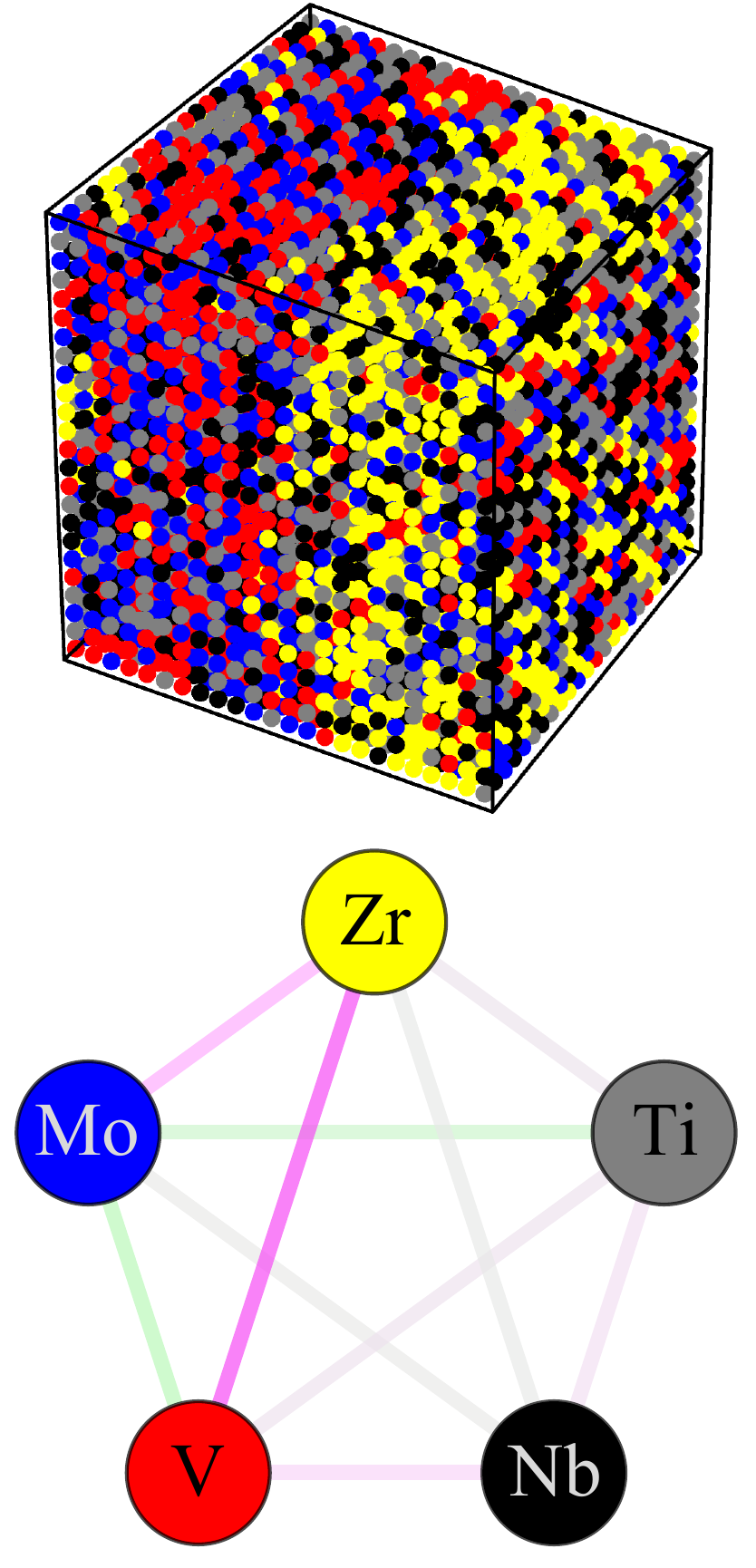}

}\hfill{}\subfloat[$T\sim2900\protect\K$ \label{fig:equicomp-high-T}]{\includegraphics[scale=0.35]{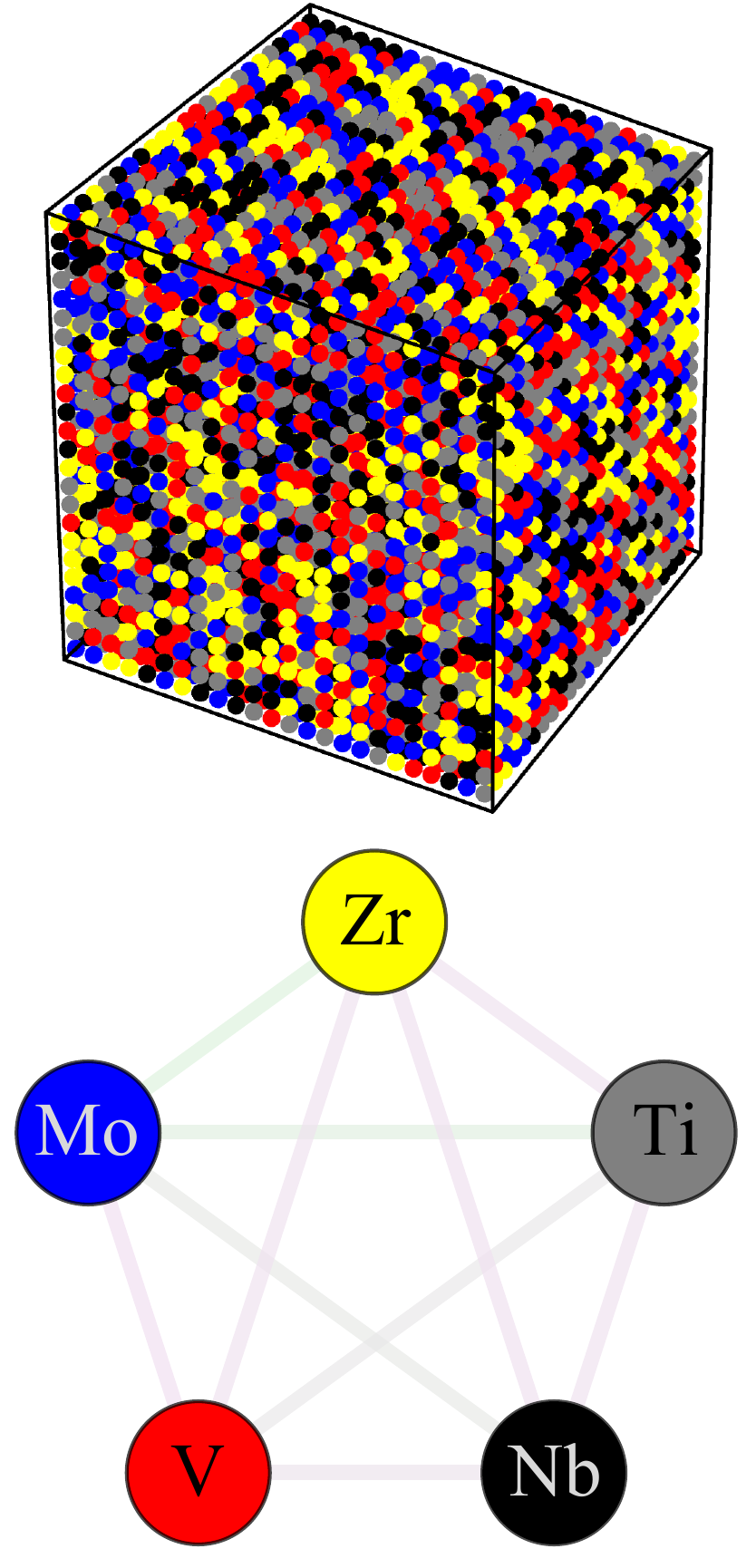}

}\hfill{}

\caption{The microstructures of equicompositional MoVNbTiZr. (a) The temperature
dependence of the ten short-range orders characterizing the microstructures.
As the solid solution cools, Zr segregates from Mo and V, while Mo
and V tend to cluster. Nb and Ti are generally evenly dispersed throughout
the system. (b-d) Snapshots of the microstructures at three representative
temperatures. Each microstructure is characterized by a graphical
representation of the SROs, where the vertices represent the five
elements and the edges are colored based on the SROs' values, according
to the bar legend in (a).}
\end{figure*}
We begin with the MC simulations of MVNTZ at equi-composition. Fig.
\ref{fig:equicomp-SRO-vs-T} illustrates the SROs of the observed
microstructures. As the high-temperature quinary solid solution cools,
Zr most notably segregates from Mo and V, while Mo and V tend to cluster.
In both phases, Nb and Ti remain randomly dispersed. We determine
a transition temperature of $T_{c}\sim1850\K$ based on the cusps
of the Zr SROs\textcolor{black}{{} (see \ref{sec:Appendix-Computing-Tc}
for details)}. As the temperature decreases further, Zr segregation
becomes more pronounced and V also segregates from Nb and Ti, while
complex intermetallics involving all five elemental species begin
forming. 

These observations are represented in Fig. \ref{fig:equicomp-low-T}-\ref{fig:equicomp-high-T},
which provide snapshots of the microstructures at three representative
temperatures. Each snapshot is accompanied by a graphical representation
of the SROs, where the vertices represent the five elemental species
and the ten edges are colored based on the SROs' values---a darker
pink (green) edge indicates a higher segregation (clustering) tendency. 

A thermodynamic description of the microstructural evolution can thus
be summarized as such: Well above the $T_{c}$ at $2900\K$, all five
elements are randomly dispersed in a single-phase solid solution.
At $1400\K$, which is slightly below the $T_{c}$, there are two
segregated secondary phases: one Zr-rich and the other a disordered
phase enriched with Mo and V, which possess a small clustering tendency.
The remaining elements Nb and Ti are distributed rather evenly between
the two phases. These observations are consistent with experimental
measurements of samples annealed at $1673\K$ \citep{Senkov2018},
except for a third (non-bcc) Laves phase which is naturally beyond
the scope of our bcc simulations. Our simulations further predict
that, way below the $T_{c}$ at $600\K$ (probably experimentally
inaccessible due to MPEAs' sluggish dynamics), there is further segregation
to at least three intermetallic phases, such as Mo-V, Mo-Ti, and Nb-Zr.

\subsection{Extension to the multidimensional compositional space}

While experiments are restricted to a small number compositions, the
computational efficiency of CE enables a high-throughput study of
the MPEAs' multidimensional compositional space. Here, we extend our
MC simulations and SRO characterization to a total of 505 compositions,
spread across the whole compositional space relevant to MPEAs. 

Across all compositions, Zr-Mo and Zr-V segregation were observed
as predominant. As expected, $T_{c}$ is highly composition-dependent,
which is modeled using a simple-to-use quadratic polynomial. With
respect to the concentrations, $\boldsymbol{x}=\left(\begin{array}{cccc}
x_{\text{Mo}} & x_{\text{V}} & x_{\text{Nb}} & x_{\text{Ti}}\end{array}\right)^{T}$:

\begin{align}
 & T_{c}/\left(1.16\times10^{4}\K\right)\label{eq:fitted-Tc}\\
 & \hfill=-0.25+\left(\begin{array}{c}
1.3\\
1.8\\
1.0\\
0.7
\end{array}\right)\boldsymbol{x}-\boldsymbol{x}^{T}\left(\begin{array}{cccc}
1.1 & 0.9 & 0.8 & 0.6\\
0.9 & 1.3 & 1.1 & 1.0\\
0.8 & 1.1 & 0.7 & 0.6\\
0.6 & 1.0 & 0.6 & 0.4
\end{array}\right)\boldsymbol{x},\nonumber 
\end{align}
where the coefficient matrix has been symmetrized. The rms error of
the fit is about $128\K$, similar to the temperature step size in
the MC simulations. Detailed in Fig. \ref{fig:fitted-Tc-vs-MC-Tc},
the equation is most reliable for $T_{c}\gtrsim1400\K$, a temperature
range which includes typical annealing temperatures in experiments.
Furthermore, fitting smooths out numerical uncertainties from the
MC simulations \textcolor{black}{(e.g. due to potentially insufficient
MC sampling)}. Most importantly, this fit enables rapid compositional
design by predicting solid solution formation across the wide concentration
ranges of MPEAs. 

\subsection{Validation from the experimental data}

\begin{figure*}
\begin{centering}
\includegraphics[scale=0.5]{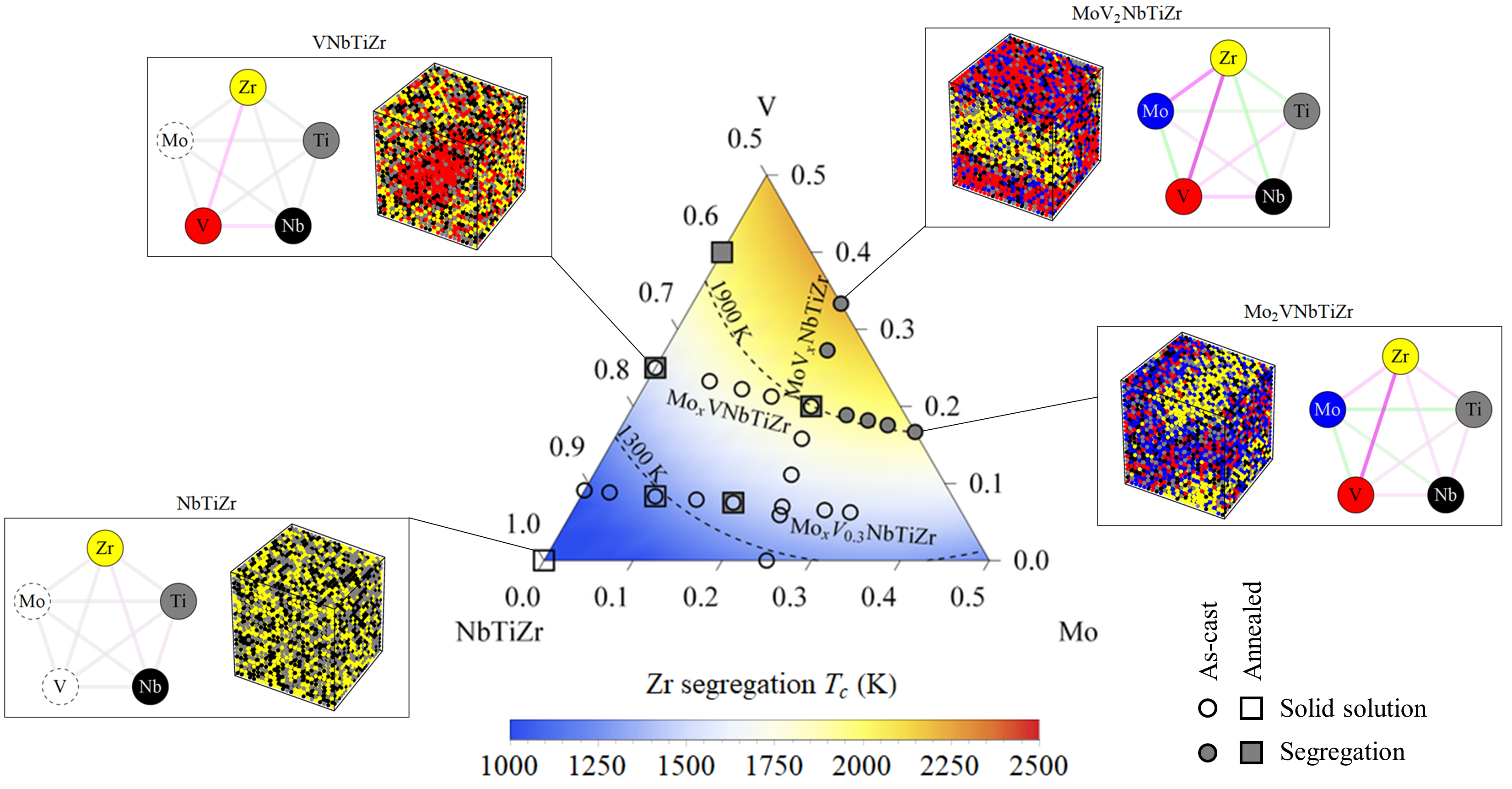}
\par\end{centering}
\caption{Experimental validation of $\text{Mo}_{x}\text{V}_{y}\text{(NbTiZr)}_{1-x-y}$
microstructures for $0\protect\leq x+y\protect\leq0.5$. A ternary
plot of the $T_{c}$ from first-principles is overlaid with existing
experimental data based on as-cast \citep{Zhang2012,Wu2015} and annealed
\citep{Senkov2013a,Wu2015,Butler2017,Senkov2018} samples. Circles
(squares) indicate compositions studied using as-cast (annealed) samples.
An unfilled (filled) symbol represents the observation of a solid
solution (segregated phase). The contour lines at $1900\protect\K$
($1300\protect\K$) guide the eye at dividing single phases from multiphases
for as-cast (annealed) samples. Both first-principles and experimental
observations agree that higher concentrations of Mo and V promote
phase segregation. The microstructures at four representative compositions
are shown for $T^{*}\sim1400\protect\K$. Each microstructure is accompanied
by a graphical representation of the SROs, where the vertices represent
the five species and the edges are colored based on values of the
SROs---a darker pink (green) edge indicates a stronger tendency to
segregate (cluster). \label{fig:Mo_xV_yNbTIZr}}
\end{figure*}
First, we focus on $\text{Mo}_{x}\text{V}_{y}\text{NbTiZr}$, where
experimental data are available to benchmark our predictions. For
as-cast samples \citep{Zhang2012,Wu2015}, $\text{Mo}_{x}\text{V}_{y}\text{NbTiZr}$
is a single-phase solid solution when either the Mo- or V-content
is low. When $x>1$ and $y=1$, the MPEA segregates into a Mo-rich,
Zr-deficient dendritic phase with Zr-rich precipitates that are deficient
in Mo and Nb. When $y=1$ and $x\geq1.5$, we similarly have Zr segregation.
For samples annealed at $1673\K$ \citep{Senkov2018}, NbTiZr is a
single-phase solid solution while VNbTiZr consists of V-rich precipitates
surrounded by Zr-rich regions. The microstructure of quinary MoVNbTiZr
is more complex: it consists of Zr-deficient dendrites surrounded
by interdendritic regions rich in Zr and Ti but deficient in Mo and
V, as well as a $(\text{Mo},\text{V})_{2}\text{Zr}$ C15 Laves phase.
Similarly, samples of $\text{V}\text{NbTiZr}$ and $\text{V}_{2}\text{NbTiZr}$
annealed at $1473\K$ contain V-rich, Zr-deficient regions and Zr-rich,
V-deficient regions \citep{Senkov2013a,Butler2017}. Finally, for
samples annealed at 1273 K \citep{Wu2015}, $\text{Mo}_{0.3}\text{V}_{0.3}\text{NbTiZr}$
remains as a single-phase solid solution, but $\text{Mo}_{0.7}\text{V}_{0.3}\text{NbTiZr}$
segregates into two bcc phases and a C15 Laves phase. 

We summarize in Fig. \ref{fig:Mo_xV_yNbTIZr} these experimental observations
of Zr segregation. Across compositions, the segregation is strongly
affected by Mo and V content. Among samples of the same type (as-cast
or annealed), phase segregation occurs at higher concentrations of
Mo and V. 

To compare with the experimental findings, we overlay our first-principles-derived
$T_{c}$ in the form of a contour plot using Eq. \ref{eq:fitted-Tc},
where a lower $T_{c}$ implies a higher tendency for the formation
of a single solid-solution phase. As shown in Fig. \ref{fig:Mo_xV_yNbTIZr},
the predicted $T_{c}$ increases with the concentration of V, in support
of experimental findings. This dependence can be traced back to Eq.
\ref{eq:fitted-Tc}, where the linear term in V-concentration is the
largest contribution. The agreement with experiments is accentuated
by the observation that contour lines of the $T_{c}$ neatly demarcate
the solid-solution samples from the phase-segregated samples, i.e.,
the solid-solution (phase-segregated) samples lie in the bluer (redder)
regions of the contour plot.

Fig. \ref{fig:Mo_xV_yNbTIZr} also showcases our MC-simulated microstructures
of a few representative systems at $T^{*}\sim1400\K$, which is close
to the experimental annealing temperatures ( 1273--$1673\K$) in
Ref. \citep{Senkov2013a,Wu2015,Butler2017,Senkov2018}. For compositions
with $T_{c}>T^{*}$, the SRO reflect the segregation of Zr from Mo
and V, similar to that at equicomposition. Across different compositions,
segregation is more pronounced for higher $T_{c}$.

\begin{figure*}[t]
\begin{centering}
\subfloat[]{\includegraphics[scale=0.4]{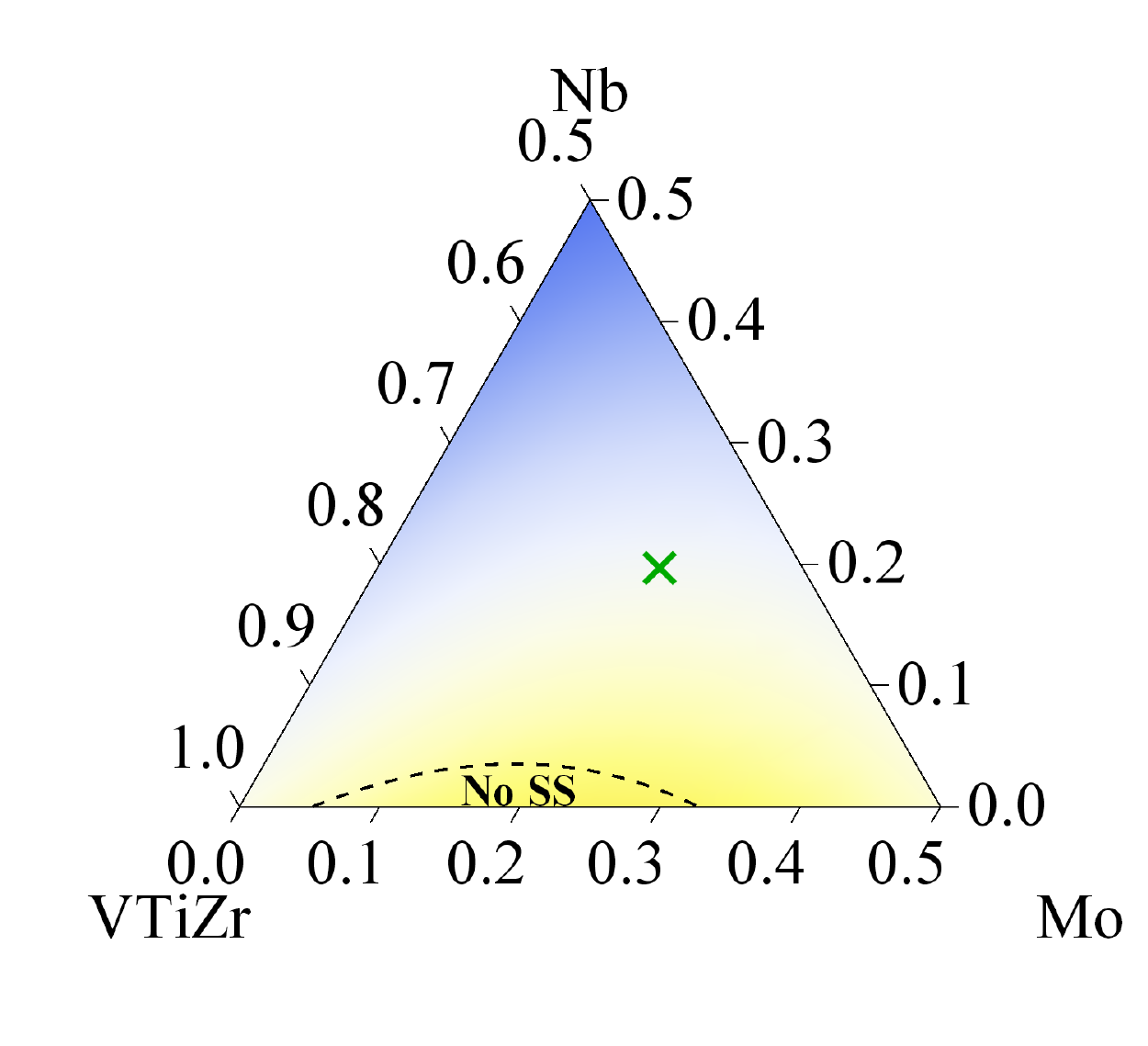}

}\subfloat[]{\includegraphics[scale=0.4]{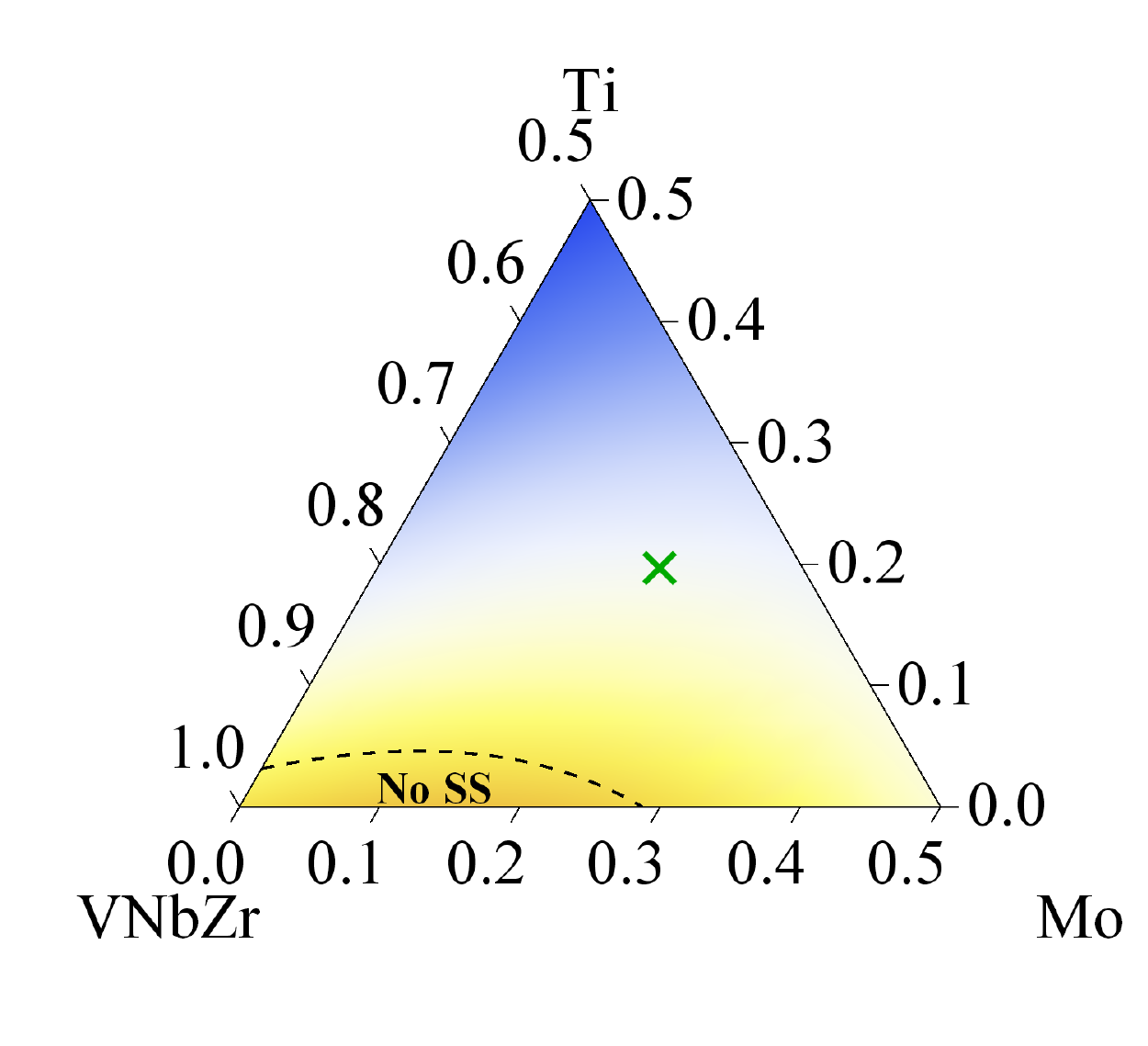}

}\subfloat[]{\includegraphics[scale=0.4]{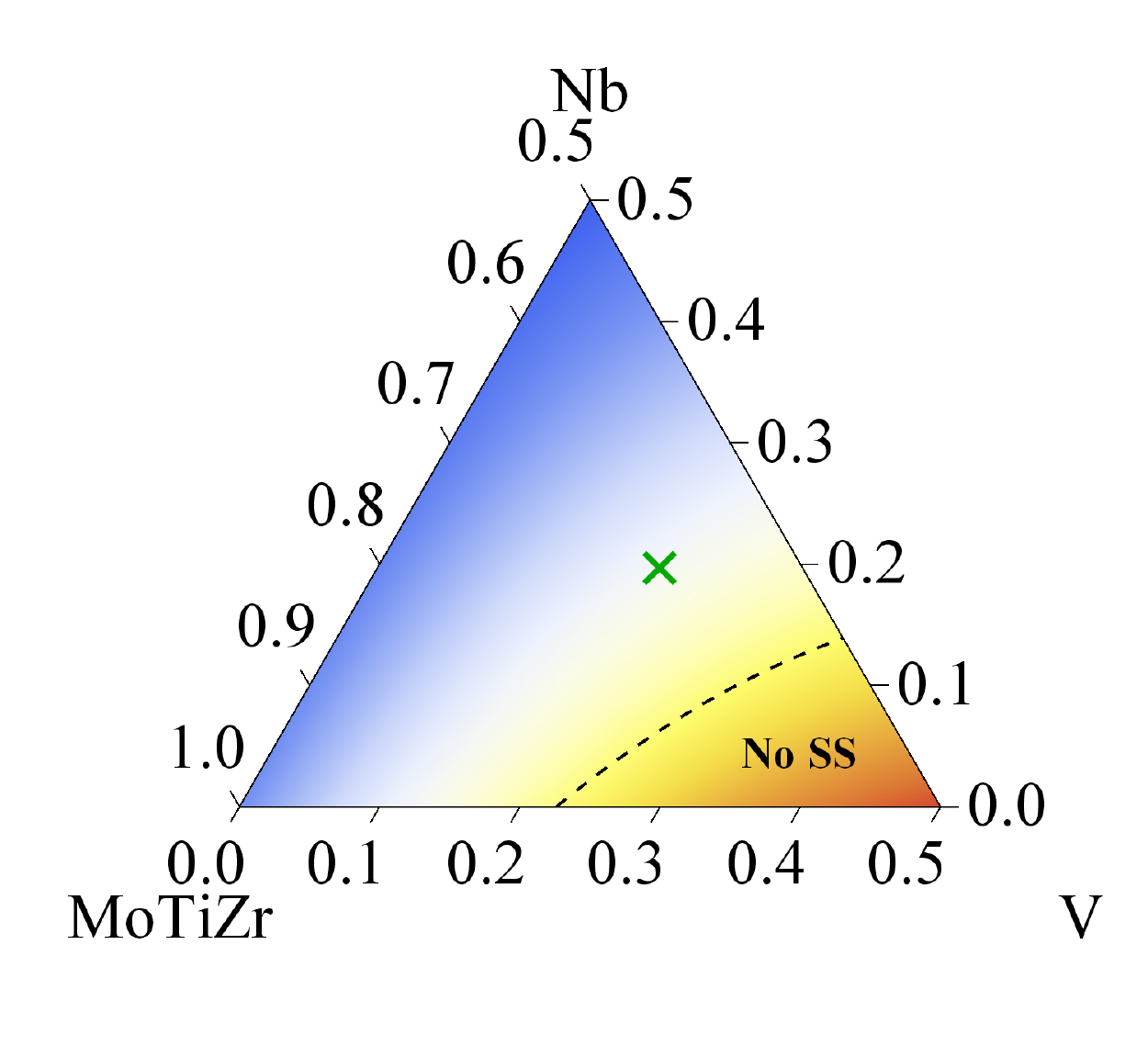}

}
\par\end{centering}
\begin{centering}
\subfloat[]{\includegraphics[scale=0.4]{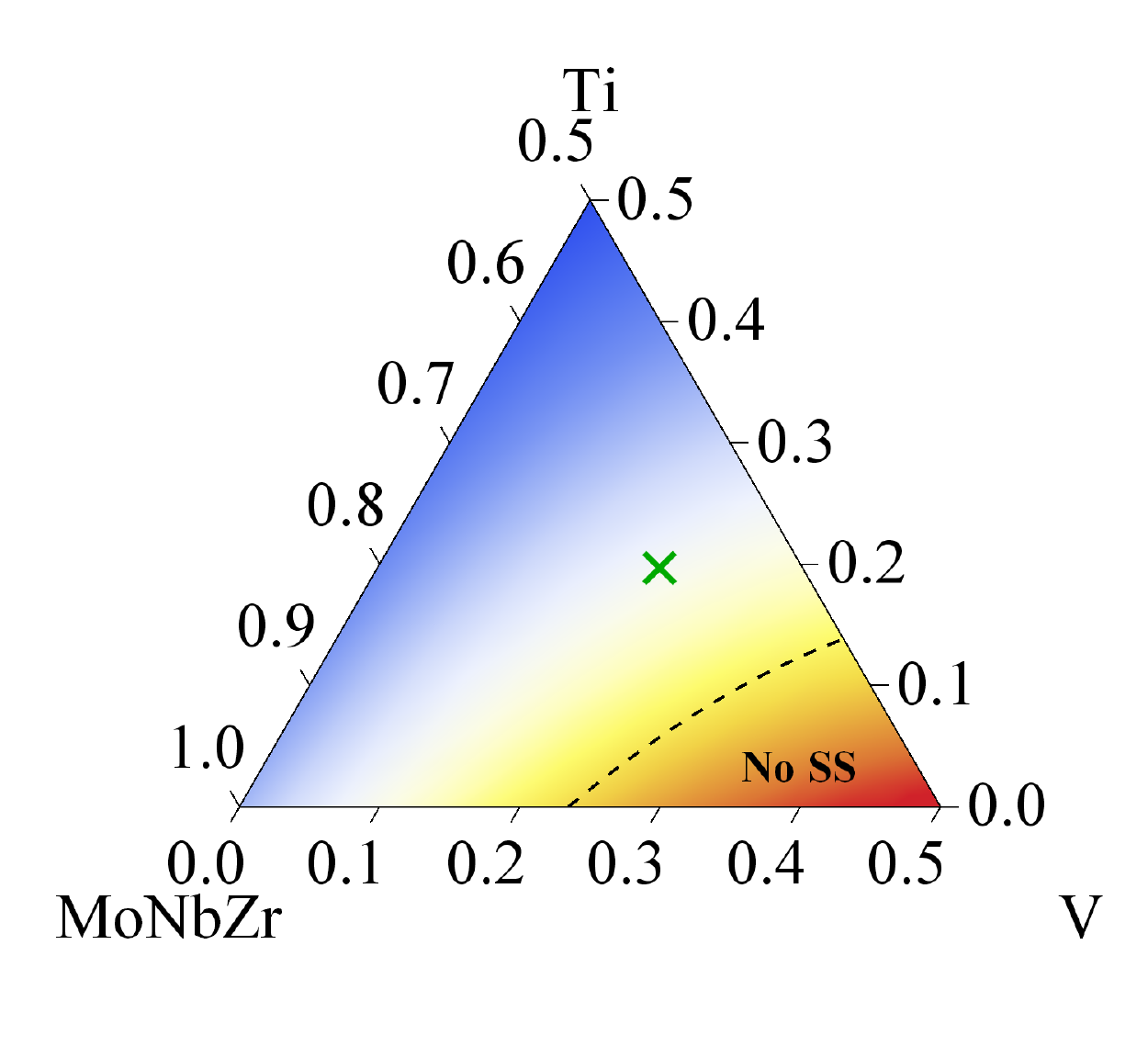}

}\subfloat[]{\includegraphics[scale=0.4]{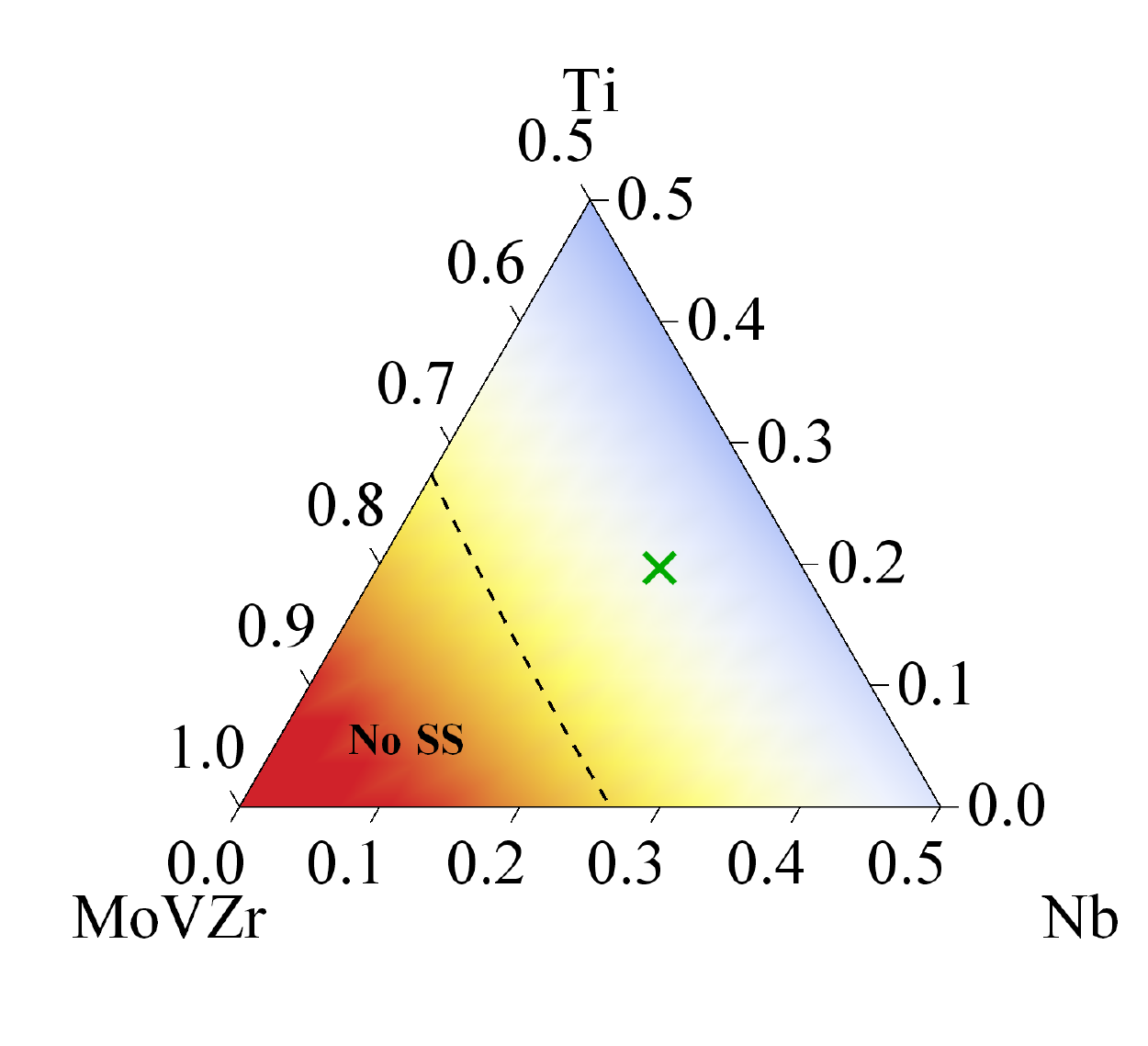}

}\subfloat[]{\includegraphics[scale=0.4]{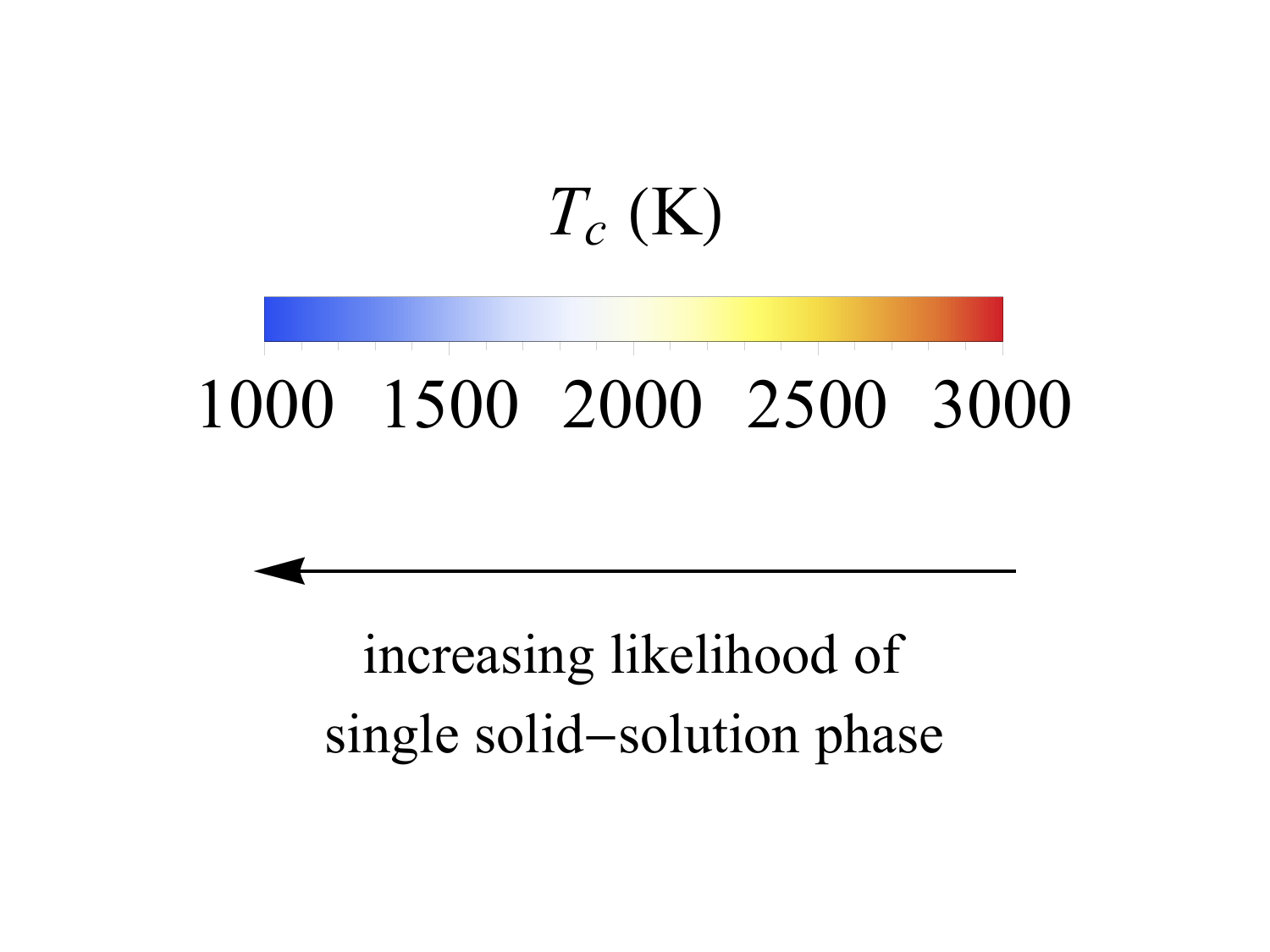}

}
\par\end{centering}
\caption{Stability of the solid solution phase in MVNTZ across compositions.
Ternary plots of the $T_{c}$ for (a) $\text{Mo}_{x}\text{V}\text{Nb}_{y}\text{Ti}\text{Zr}$,
(b) $\text{Mo}_{x}\text{V}\text{Nb}\text{Ti}_{y}\text{Zr}$, (c) $\text{Mo}\text{V}_{x}\text{Nb}_{y}\text{Ti}\text{Zr}$,
(d) $\text{Mo}\text{V}_{x}\text{Nb}\text{Ti}_{y}\text{Zr}$, and (e)
$\text{Mo}\text{V}\text{Nb}_{x}\text{Ti}_{y}\text{Zr}$, for $0\protect\leq x+y\protect\leq0.5$,
with the legend given in (f). In each plot, the dashed line is given
by $T_{c}=T_{m}$, the melting temperature. When $T_{c}>T_{m}$, a
solid solution phase is prohibited. The green cross in each plot indicates
equicomposition. \label{fig:Ternary-plot-of-Tc-vs-composition} }
\end{figure*}
Fig. \ref{fig:Ternary-plot-of-Tc-vs-composition} presents the analogous
$T_{c}$ ternary plots where the concentrations of other pairs of
elements are varied: $\text{Mo}_{x}\text{V}\text{Nb}_{y}\text{Ti}\text{Zr}$,
$\text{Mo}_{x}\text{V}\text{Nb}\text{Ti}_{y}\text{Zr}$, $\text{Mo}\text{V}_{x}\text{Nb}_{y}\text{Ti}\text{Zr}$,
\linebreak{}
 $\text{Mo}\text{V}_{x}\text{Nb}\text{Ti}_{y}\text{Zr}$, and $\text{Mo}\text{V}\text{Nb}_{x}\text{Ti}_{y}\text{Zr}$,
for $0\leq x+y\leq0.5$. In these cases, the Zr content is deliberately
substantial because of its role in the phase segregations.

Unlike $\text{Mo}_{x}\text{V}_{y}\text{Nb}\text{Ti}\text{Zr}$ in
Fig. \ref{fig:Mo_xV_yNbTIZr}, there exist limited experimental studies
of the microstructures for these compositions \citep{Mu2017}. Therefore,
our computational results here are predictions of the MVNTZ microstructures
in the unexplored regions of the compositional space: MVNTZ phase
segregates similarly to Fig. \ref{fig:Mo_xV_yNbTIZr}, with the $T_{c}$
given by Fig. \ref{fig:Ternary-plot-of-Tc-vs-composition} and Eq.
\ref{eq:fitted-Tc}. This knowledge of the temperature and compositional
dependence of the microstructure can be used to tune the mechanical
properties of the MPEAs using different thermal histories \citep{He2016},
through various strengthening mechanisms \citep{Basu2020}.

A lower $T_{c}$ implies a more stable solid solution phase. Hence,
the bluer regions in Fig. \ref{fig:Ternary-plot-of-Tc-vs-composition}
illustrate the compositions for which a (single-phase) solid solution
is more likely to be found. This is true provided that the predicted
$T_{c}$ is less than $T_{m}$, the melting temperature. Therefore,
we have also indicated in Fig. \ref{fig:Ternary-plot-of-Tc-vs-composition}
the compositions for which $T_{c}>T_{m}$ (see \ref{sec:Appendix-Tm}).
Note that for $\text{Mo}_{x}\text{V}_{y}\text{NbTiZr}$ in Fig. \ref{fig:Mo_xV_yNbTIZr},
$T_{c}<T_{m}$ at all points in the ternary diagram. 

\textcolor{black}{Taking into account practical considerations, we
propose three specific compositions for future experimental studies.
Based on the MC simulations, each of these systems is a single-phase
solid solution at 1$400\K$, which is around the typical annealing
temperature. By favoring Ti over Mo, the MPEA $\text{V}\text{Nb}\text{Ti}_{2}\text{Zr}$
is a solid solution with reduced density ($\rho\sim6.08\ \text{g}\ \text{cm}^{-3}$
based on the rule of mixtures). Such an MPEA could be useful as lightweight
alloys in the aerospace industry \citep{Senkov2013a,Senkov2013}.
Among the five elemental species, Ti has the lowest density ($\rho_{\text{Ti}}=4.51\ \text{g}\ \text{cm}^{-3}$)
while Mo has the highest ($\rho_{\text{Mo}}=10.3\ \text{g}\ \text{cm}^{-3}$).
By omitting V, the MPEAs $\text{Mo}\text{Nb}\text{Ti}_{2}\text{Zr}$
and $\text{Mo}\text{Nb}_{2}\text{Ti}\text{Zr}$ are solid solutions
with improved biocompatibility, as V ions are known to exhibit cytotoxicity
\citep{Okazaki1998,Lopez2003}. Such MPEAs could be useful as biomedical
implants. }

\subsection{Origin of phase segregation}

\begin{figure}[t]
\subfloat[\label{fig:microstructure-no-Mo}]{\includegraphics[scale=0.3]{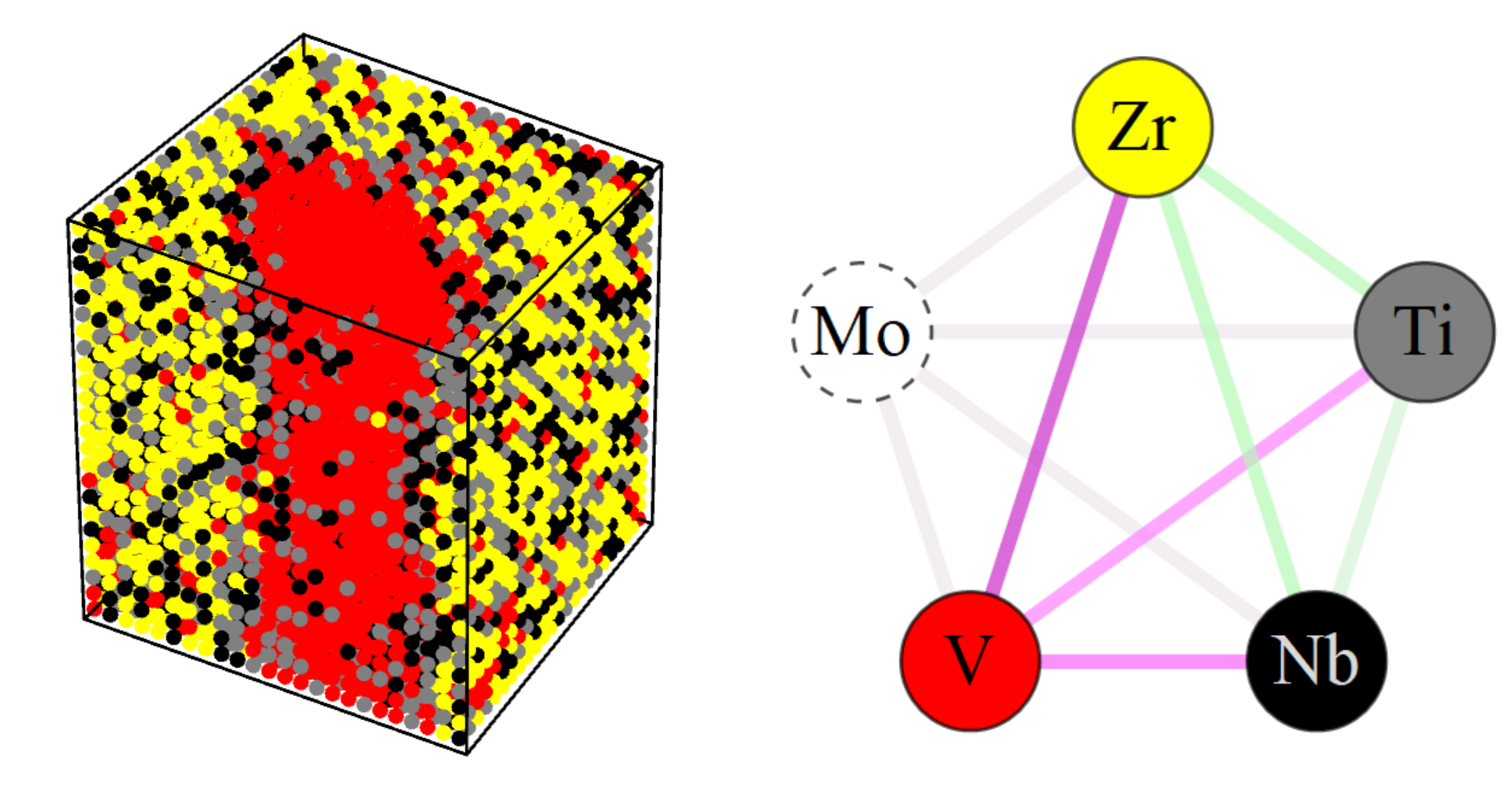}

}

\subfloat[\label{fig:microstructure-no-V}]{\includegraphics[scale=0.3]{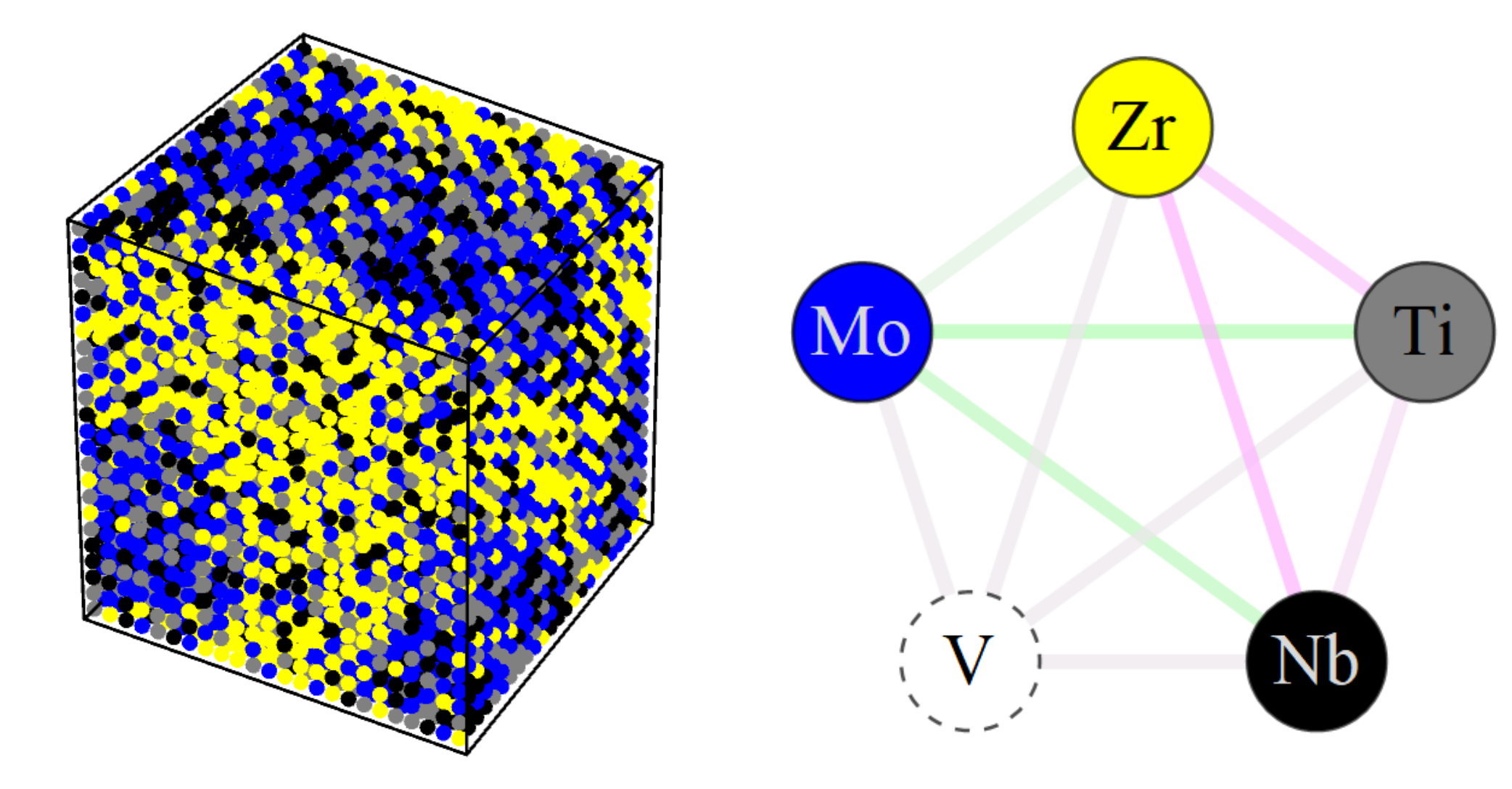}

}

\subfloat[\label{fig:microstructure-no-Zr}]{\includegraphics[scale=0.3]{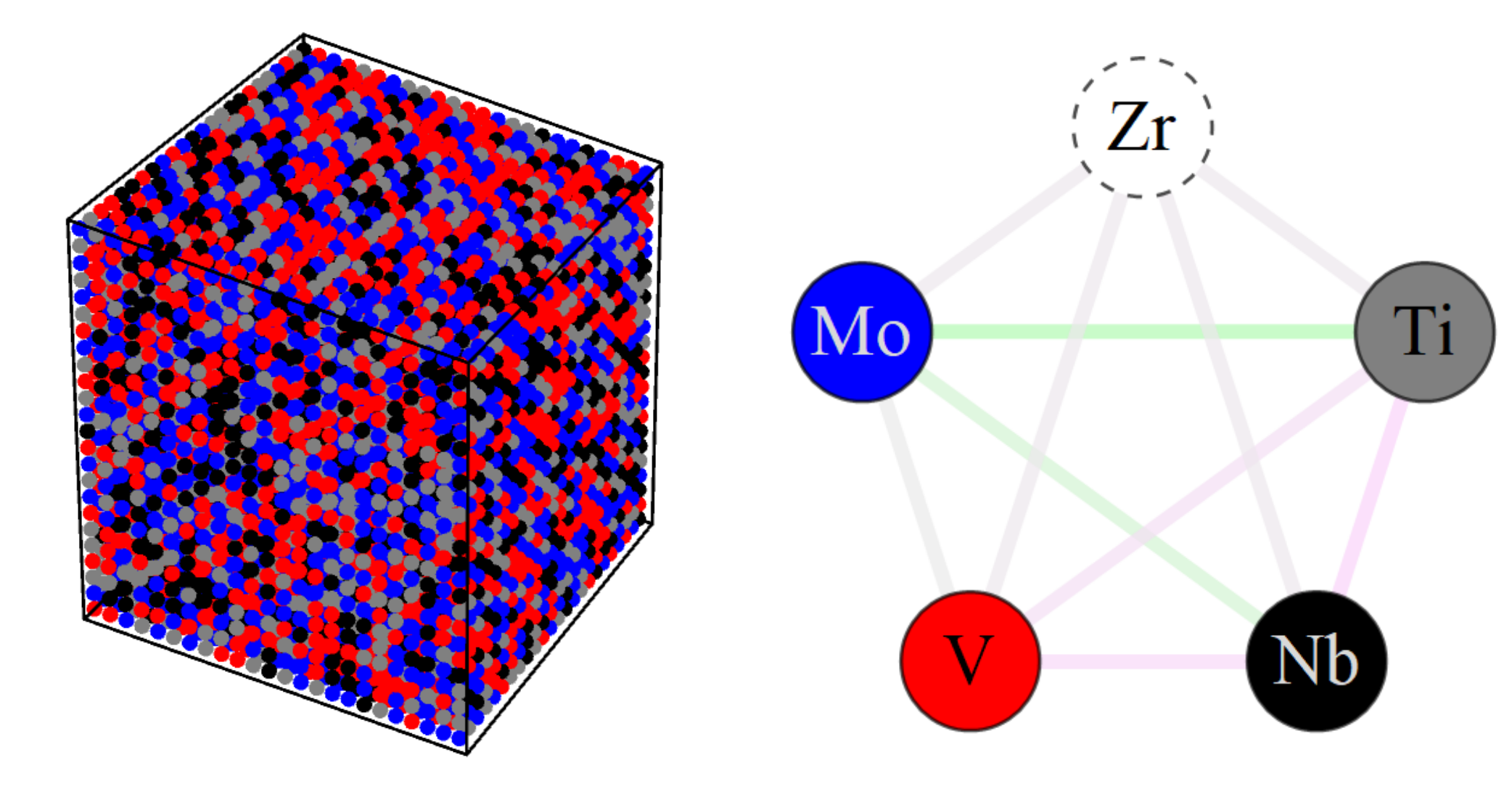}

}

\caption{Three quaternary edge cases. The microstructures and SROs of (a) $\text{V}_{3}\text{Nb}_{2}\text{Ti}_{2}\text{Zr}_{3}$,
(b) $\text{Mo}_{3}\text{Nb}_{2}\text{Ti}_{2}\text{Zr}_{3}$, and (c)
$\text{Mo}_{3}\text{V}_{3}\text{Nb}_{2}\text{Ti}_{2}$ at $1000\protect\K$.
In the graphical representations of the SROs, the vertices represent
the five species and the edges are colored based on values of the
SROs---a darker pink (green) edge indicates a stronger tendency to
segregate (cluster). The vertices of excluded elements are shown in
white with dashed borders. \label{fig:microstructures-edge-cases}}
\end{figure}
While Zr segregates from Mo and V throughout the compositional space,
one may wonder what happens when either Mo or V is absent from it.
Here, we illustrate the microstructures of two such cases (Fig. \ref{fig:microstructures-edge-cases}),
compared with that at equicomposition (Fig. \ref{fig:equicomp-mid-T}).

Without Mo, the quaternary system in Fig. \ref{fig:microstructure-no-Mo}
is best described as V segregated. While Zr-V segregation remains
predominant, Nb and Ti are now concentrated in the Zr-rich phase.
This shows that Nb and Ti prefer clustering with Zr over V, forming
a V-rich phase and a phase rich in Zr, Nb, and Ti. Experimentally,
samples of VNbTiZr and $\text{V}_{2}$NbTiZr annealed at $1473\K$
phase segregates into Zr-rich and V-rich bcc phases \citep{Senkov2013a}.
Compression deformation at $1273\K$ of the $\text{V}_{2}$NbTiZr
samples induces a more complete transformation such that the V-rich
phase also becomes Ti-deficient \citep{Senkov2013}. These observations
are consistent with the results from MC simulations. 

Without V, Fig \ref{fig:microstructure-no-V} shows that Zr now segregates
from Nb and Ti, while Mo is evenly distributed. Zr-Mo segregation
is noticeably suppressed, suggesting that Zr-Mo segregation at equicomposition
is driven by the combination of Zr-V segregation and Mo-V clustering
tendency. Furthermore, we observe that that Nb and Ti prefer clustering
with Mo over Zr. This suggests that, at equicomposition, Nb and Ti
has a tendency to cluster with Mo but segregate from V. This tendency
competes against the Mo-V clustering tendency, suppressing the segregation
of V from Nb and Ti. This suppression is well illustrated by comparing
Figs. \ref{fig:microstructure-no-Mo} and \ref{fig:microstructure-no-Zr},
which contains Mo instead of Zr. In this case, we also observe that,
without Zr, the $T_{c}$ is low ($<1000\K$), confirming Zr's role
in the predominant behavior of MVNTZ at equicomposition. This is consistent
with the observation of a solid solution in as-cast MoVNbTi \citep{Chen2014}.

To summarize, we provide four key principles governing the thermodynamics
of MVNTZ MPEAs:
\begin{lyxlist}{00.00.0000}
\item [{(1):}] V-Zr segregation is the predominant behavior. 
\item [{(2):}] Mo and V tend to cluster. This and (1) together promote
Mo-Zr segregation. 
\item [{(3):}] Nb and Ti have a hierarchical clustering tendency, preferring
Mo over Zr, and Zr over V. 
\item [{(4):}] The competition between (2) and (3) moderates the tendency
for V-Nb and V-Ti segregation and Mo-V clustering.
\end{lyxlist}
In general, the precise behaviors at any composition is a fine balance
between the tendency for certain elemental species to cluster or segregate.
These principles help elucidate the origin of phase segregation in
the quinary system, providing a deeper understanding of the microstructural
behaviors of MPEAs. 

\section{Discussion}

\begin{figure}[t]
\subfloat[]{\includegraphics[scale=0.35]{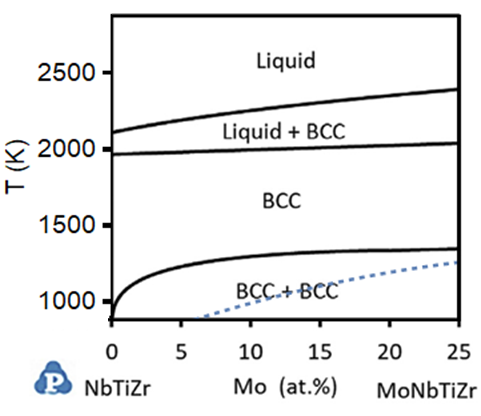}

}

\subfloat[]{\includegraphics[scale=0.35]{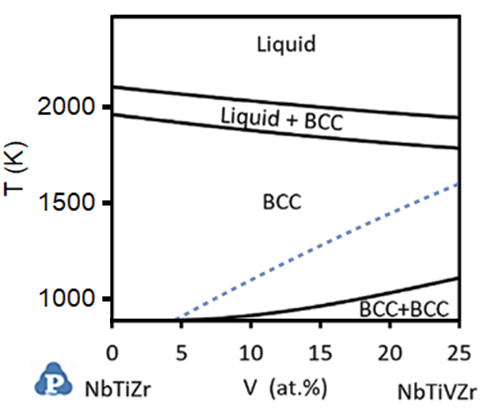}

}

\caption{A comparison of the CALPHAD phase diagrams (solid lines) from Ref.
\citep{Senkov2019} with our CE-based $T_{c}$ (dashed lines), for
(a) $\text{Mo}_{x}\left(\text{NbTiZr}\right)_{1-x}$ and (b) $\text{V}_{x}\left(\text{NbTiZr}\right)_{1-x}$.
\label{fig:CALPHAD-comparison}}
\end{figure}
In Fig. \ref{fig:CALPHAD-comparison}, we compare our $T_{c}$ from
Eq. \ref{eq:fitted-Tc} with the results based on CALPHAD from Ref.
\citep{Senkov2019}. Both sets of results are consistent for practical
purposes---experiments are unlikely to favor one set of computational
results over the other given the slow kinetics in the MPEA, especially
at lower temperatures.\textcolor{black}{{} This is true except for VNbTiZr,
where CALPHAD predicts a much lower $T_{c}$ ($\sim1100\K$) than
our first-principles-derived results ($\sim1600\K$). Existing experimental
data shows that VNbTiZr phase segregates even for samples annealed
between $1473$--$1673\K$ \citep{Senkov2013a,Butler2017,Senkov2018},
contradicting the low $T_{c}$ that CALPHAD predicts. As noted in
Ref. \citep{Senkov2019,Miracle2019}, reliable CALPHAD results for
MPEAs require a complete thermodynamic database---extrapolations
from incomplete data of the constituent binary/ternary alloys could
produce incorrect results. Predicting the phases of MPEAs using CALPHAD
remains an active field of research \citep{Li2020b,Zeng2021}. In
contrast, our model goes beyond the extrapolation of binary or ternary
data, since the first-principles training data for our model include
quaternary and quinary structures as well. The importance of these
higher order contributions is further highlighted by the sizable ECIs
(shown in Fig. \ref{fig:glasso-ECIs}) for clusters containing four
or more different elemental species. Therefore, this comparison for
VNbTiZr showcases an example in which the MC-derived $T_{c}$, built
upon first-principles data computed over all relevant MPEA compositions
(from binary to quinary), is more consistent with experiments than
CALPHAD is. }

In Fig. \ref{fig:Ternary-plot-of-Tc-vs-composition}, we identified
regions within the composition space for which solid solution formation
is prohibited. The accuracy of these regions undoubtedly depends on
the $T_{m}$ estimates used. In fact, a more accurate treatment would
involve the temperature range over which solidification takes place.
Doing so would enlarge, in Fig. \ref{fig:Ternary-plot-of-Tc-vs-composition},
the bounded regions that prohibit solid solution formation. 

Our DFT and MC calculations are based on a bcc lattice because each
of the five elements of MVNTZ forms a bcc lattice in the solid solution
phase. This is true even for Ti and Zr, which form hcp structures
at low temperatures. Furthermore, existing experimental data of MVNTZ
also show a predominant bcc phase \citep{Zhang2012,Senkov2013a,Wu2015,Butler2017,Senkov2018}.
However, given the limited understanding of MPEAs, phases with other
lattice structures, such as the Laves phase, could also exist. Reproducing
such phases is possible in principle, with separate sets of DFT and
MC calculations for different lattices. 

\section{Conclusions}

We have presented a high-throughput first-principles study of refractory
MPEA Mo-V-Nb-Ti-Zr. Based on the formation energies of a set of highly
representative structures, we constructed a cluster expansion model
that enables high-throughput Monte Carlo simulations. Across the entire
compositional design space, we predicted that the high-temperature
solid solution cools into two segregated secondary phases: one Zr-rich
and the other enriched with Mo and V, while Nb and Ti remain evenly
distributed. By characterizing the microstructures using short-range
order, we identify the predominant driving force to be Zr-V segregation,
which together with the clustering tendency of Mo and V, also leads
to Mo-Zr segregation. Quantitatively, we encapsulated the results
of our high-throughput study in a practical expression for predicting
the $T_{c}$ from the composition. We showed that the $T_{c}$, and
hence the tendency for phase segregation, most significantly increases
with V content, and we identified compositions for which solid solution
formation is prohibited. Within a consistent framework, our results
reproduce the phases observed in numerous sets of experiments from
literature. Our results provide the highly desired insights into the
MPEA's microstructures, guiding future experiments towards the appropriate
regions within the composition space when designing MPEAs with superior
mechanical properties.

\section*{Acknowledgments}

The authors thank the Advanced Manufacturing and Engineering Young
Individual Research Grant (AME YIRG) of Agency for Science, Technology
and Research (A{*}STAR) (A1884c0016) for financial support. The DFT
computations in this article were performed on the resources of the
National Supercomputing Centre, Singapore \linebreak{}
(https://www.nscc.sg).

\appendix

\section{DFT calculations \label{sec:Appendix-DFT}}

The energies of the CE training structures are calculated based on
density functional theory (DFT) with the Vienna Ab initio Simulation
Package (VASP) \citep{Kresse1996,Kresse1996a}. We use the Perdew,
Burke, and Ernzerhof exchange correlation based on the generalized
gradient approximation \citep{Perdew1996,Perdew1997}. The projector
augmented-wave potentials are used with the outer $p$ semicore states
included in the valence states \citep{Blochl1994,Kresse1999}. Planewave
cutoffs are set to $520\eV$. For each structure, two sets of constrained
relaxation are performed. In the initial relaxation, only the cell
volume is relaxed, keeping the atom positions and cell shape fixed
to that of a bcc lattice (ISIF = 7). From this semi-relaxed structure,
we further relaxed the atomic positions (ISIF = 2); this results in
structures with cubic unit cells but with atom positions shifted from
their ideal bcc positions. The resulting structures are thus representative
of MPEAs/HEAs at moderate to high temperatures, where the overall
structure is observed to have a cubic symmetry but with the possibility
of local lattice distortions. The energy reduction during the second
stage of relaxation gives the structural distortion energy that quantifies
the degree of distortion in Fig. \ref{fig:CE-vs-DFT-energies}. Our
convergence criterion is based on keeping the Hellmann-Feynman force
on each atom below $0.015\eV/\text{\AA}$. Calculations are not spin
polarized as Mo, Nb, V, Ti, and Zr are not known to be strongly magnetic.
The $k$-point mesh is generated using a $\Gamma$ grid and density
of $200\ \text{\AA}^{-3}$.

\section{Cluster expansion \label{sec:Appendix-CE}}

Based on the generalized Ising model, CE expands the configurational
energy $E\left(\sigma\right)$ of an alloy structure, $\sigma$, in
terms of atomic clusters $\alpha$, where the cluster correlation
functions $\Phi_{\alpha}\left(\sigma\right)$ serve as the basis set
and the effective cluster interactions (ECIs) $V_{\alpha}$ serve
as the coefficients: 
\begin{equation}
E\left(\sigma\right)=\sum_{\alpha}\Phi_{\alpha}\left(\sigma\right)V_{\alpha}.
\end{equation}
By fitting to first-principles energies, the ECIs between various
elemental species in the alloy can be determined.

The 10265 training structures of the CE are selected from a pool of
derivative structures, systematically generated up to a six-atom unit
cell \citep{Hart2008,Hart2012}. Shown in Fig. \ref{fig:CE-structures-barchart},
the structures contain between two to five elemental species. Our
CE-derived models are therefore built upon data spread across a large
compositional space, including up to quinary data. This is unlike
CALPHAD, whose model are based on binary or ternary data only.

In our CE, we treat V, Nb, Ti, and Zr as the independent species,
while Mo is treated as dependent. As such, only clusters formed by
V, Nb, Ti, and Zr atoms are required. In the bcc lattice, we consider
up to the 12th-nearest-neighbor (12NN) pairs, 5NN triplets, and 3NN
four-body to six-body clusters. These correspond to an initial pool
of 2911 symmetrically distinct clusters, consisting of 120 pairs,
596 triplets, 565 four-body clusters, 1080 five-body clusters, and
550 six-body clusters. 

Our fitting procedure follows our earlier work in Ref. \citep{Leong2019}.
Using the DFT formation energies of the training structures, we use
group lasso to select a properly truncated CE set from the initial
2911 distinct clusters. Fivefold cross-validation is used for selecting
the optimal hyperparameter with the one-standard-error rule. In the
trained CE, 2175 out of the initial 2911 clusters are selected. 

\section{Melting temperature \label{sec:Appendix-Tm}}

The melting temperature $T_{m}$ in Fig. \ref{fig:Ternary-plot-of-Tc-vs-composition}
is approximated by the simple weighted average of the compositions:
$T_{m}=\sum_{i}x_{i}T_{m,i}$, where the elemental melting temperatures
in ascending order are $T_{m,Ti}=1941\K$, $T_{m,Zr}=2128\K$, $T_{m,V}=2183\K$,
$T_{m,Nb}=2750\K$, and $T_{m,Mo}=2896\K$.

\section{Structural distortion in cluster expansion \label{sec:Appendix-distortion}}

\setcounter{figure}{0}

\begin{figure}
\includegraphics[scale=0.5]{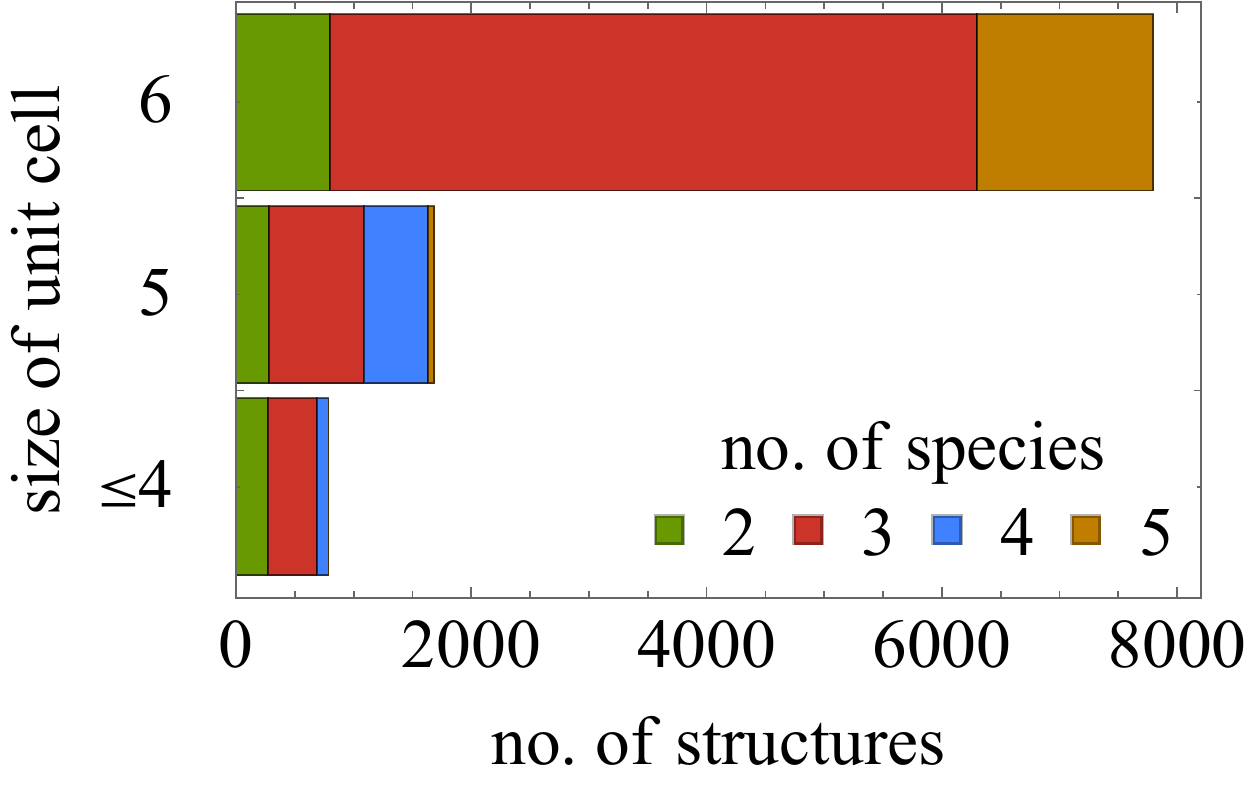}

\caption{Among the 10265 derivative structures used for training the CE models,
the bar chart shows the distribution of unit cell sizes and the number
of elemental species. \label{fig:CE-structures-barchart}}
\end{figure}
\begin{figure}
\subfloat[\label{fig:CE-error-vs-distortion}]{\includegraphics[scale=0.4]{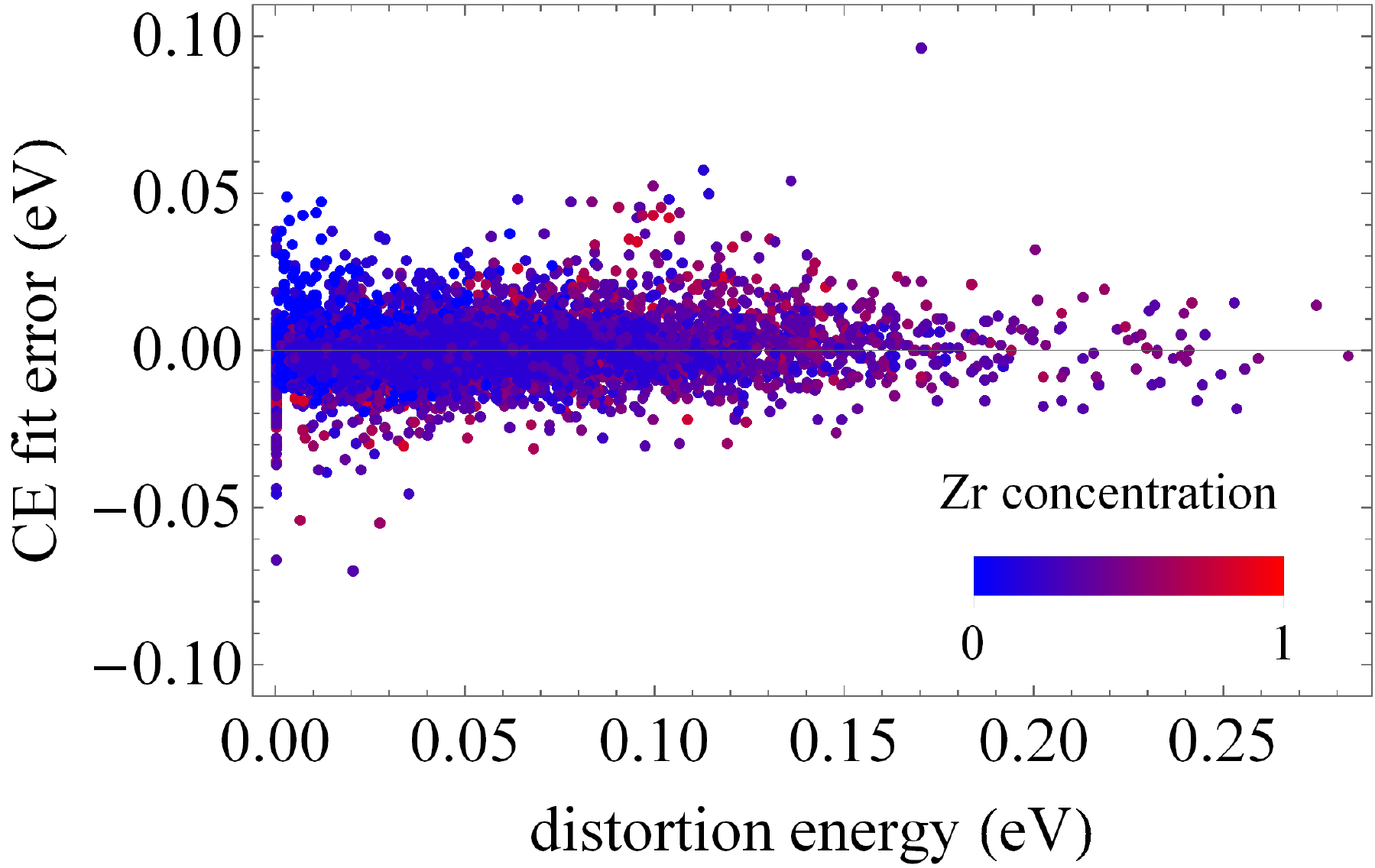}

}

\subfloat[\label{fig:distortion-correlation}]{\includegraphics[scale=0.5]{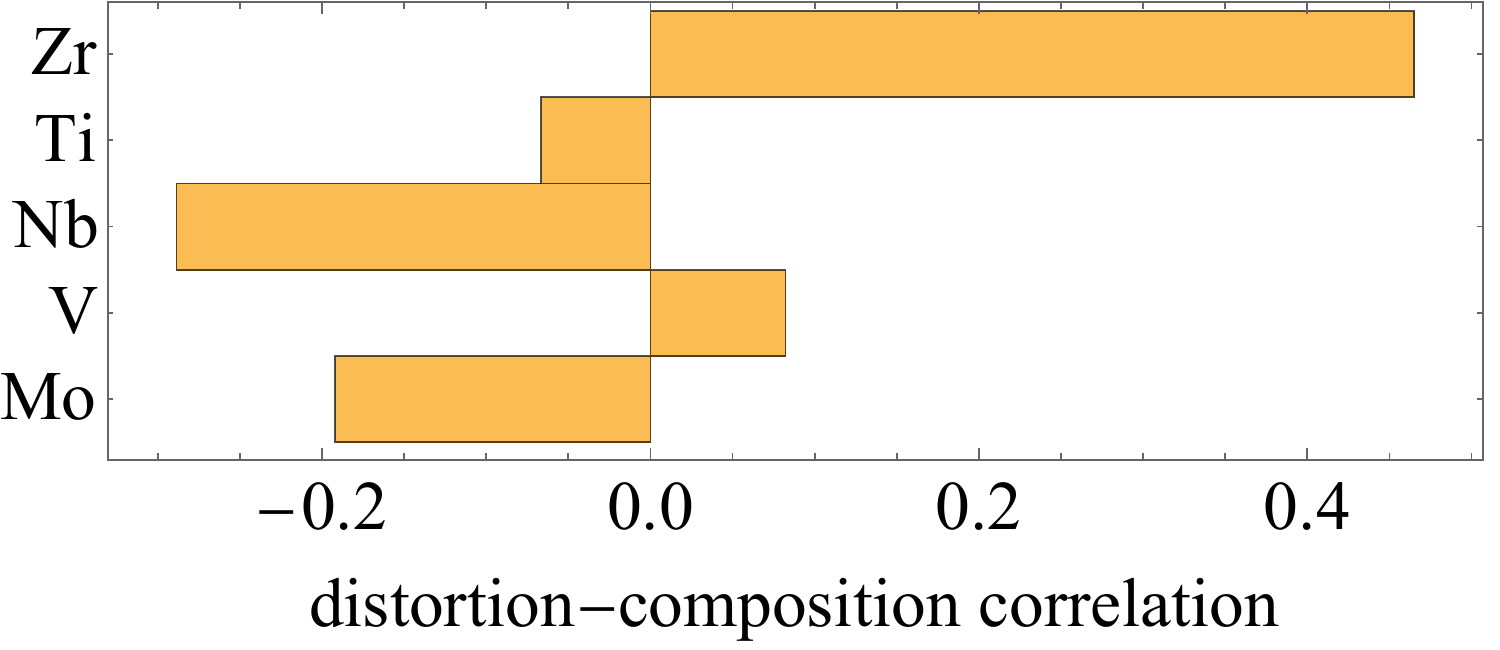}

}

\caption{(a) A plot of the CE fitting error against the distortion energy for
the 10265 training structures. Bluer (Redder) points correspond to
structures with lower (higher) Zr content. (b) The correlation between
distortion energy and concentration of each of the five elements.
Here, correlation ranges from -1 to 1. Structures with higher Zr content
tend to have more structural distortion. }
\end{figure}

For the 10265 derivative structures detailed in Figure \ref{fig:CE-structures-barchart},
we quantify the degree of structural distortion in DFT by the distortion
energy. Figure \ref{fig:CE-error-vs-distortion} shows that our CE
fit is accurate even for structures with large distortions, implying
that the CE has captured the effects of distortion. Also, we observe
that structures with higher Zr content tends to have larger distortion.
We quantify this by calculating the correlation (Pearson correlation
coefficient) between the distortion energy and the concentration of
each elements. Figure \ref{fig:distortion-correlation} shows that
Zr, among the five elements, contributes most to distortion. 

\begin{figure}
\subfloat[\label{fig:ECIs-comparison-distortion}]{\includegraphics[scale=0.4]{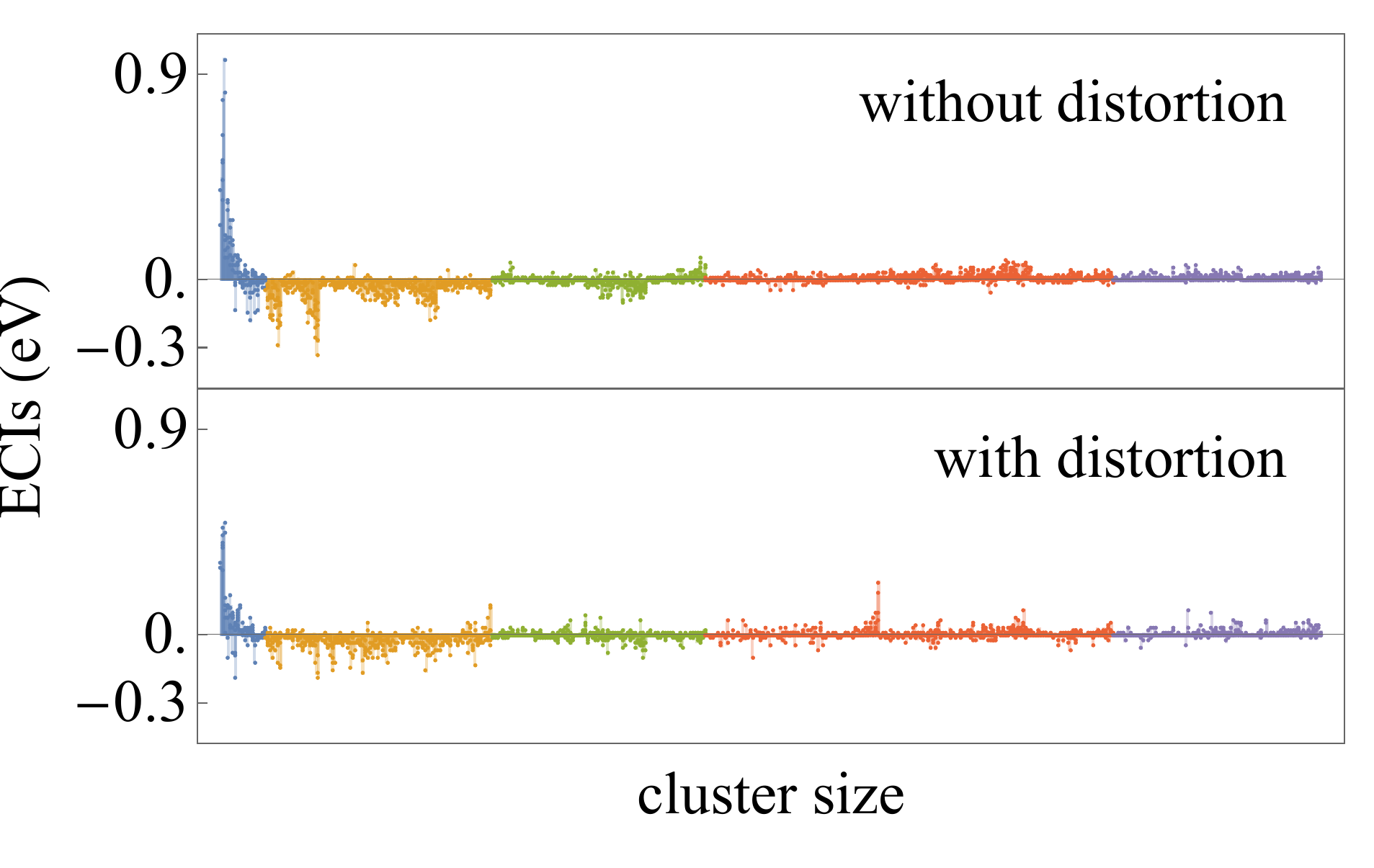}

}

\subfloat[\label{fig:CE-vs-DFT-energies-unrelaxed}]{\includegraphics[scale=0.4]{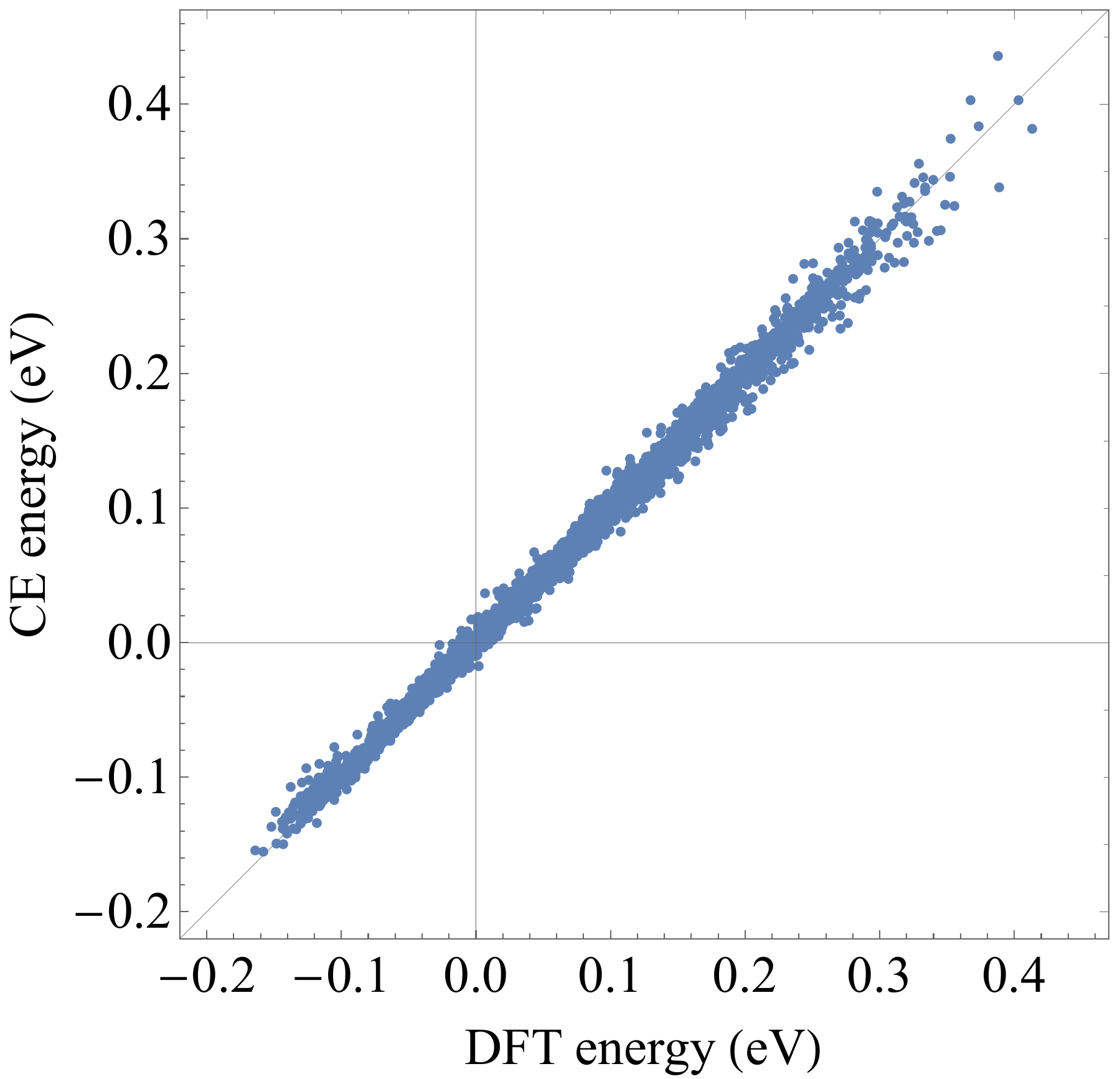}

}

\caption{(a) Comparing the two sets of 2911 ECIs from CEs based on structures
(bottom) with and (top) without structural distortion. Pairs, triplets,
quadruplets, five-bodies, and six-bodies are colored blue, orange,
green, red, and purple, respectively. (b) A plot of the predicted
CE energies against the DFT energies for the training structures without
distortion. The cross-validation score is $5.4\protect\meV$. \label{fig:unrelaxed-plots}}
\end{figure}

To elucidate the effects of structural distortion, we train another
CE model with structures without distortion, where the atomic positions
are unrelaxed and remain in their original ideal bcc lattice positions.
Figure \ref{fig:ECIs-comparison-distortion} compares the resulting
ECIs of CEs with and without distortion. Without distortion, ECIs
are well behaved and without large fluctuations in higher-order ECIs
(triplets and beyond). The effects of distortion are captured by shifting
contributions from pair ECIs to the ECIs of the relevant higher-order
clusters. Finally, Figure \ref{fig:CE-vs-DFT-energies-unrelaxed}
shows that without distortion, the CV score is $5.4\meV$. With distortion,
the slightly higher CV score quantifies the extent CE captures the
effects of distortion: CE captures distortion partially, but well.


\section{\textcolor{black}{Computing the $T_{c}$ \label{sec:Appendix-Computing-Tc}}}

\textcolor{black}{\setcounter{figure}{0}}

\textcolor{black}{The $T_{c}$'s are computed from the cusps of the
SROs of Zr with respect to all the other elements. Zr SROs are used
because Zr segregation is the predominant behavior in the system.
As the solid solution phase cools, Zr is the first species to undergo
a transition, as indicated by the cusps in the SROs. For any elemental
pair, the SRO\textquoteright s cusp is defined as the temperature
at which the magnitude of the SRO\textquoteright s second derivative
is maximum: $T_{\text{cusp}}=\amax_{T}\left|d^{2}\text{SRO}/dT^{2}\right|$,
with the derivatives computed using finite differences. Among these
SROs, the highest-temperature cusp determines the $T_{c}$. }

\section{Fitting the $T_{c}$ \label{sec:Appendix-Tc-fit}}

\setcounter{figure}{0}
\begin{center}
\begin{figure}
\includegraphics[scale=0.4]{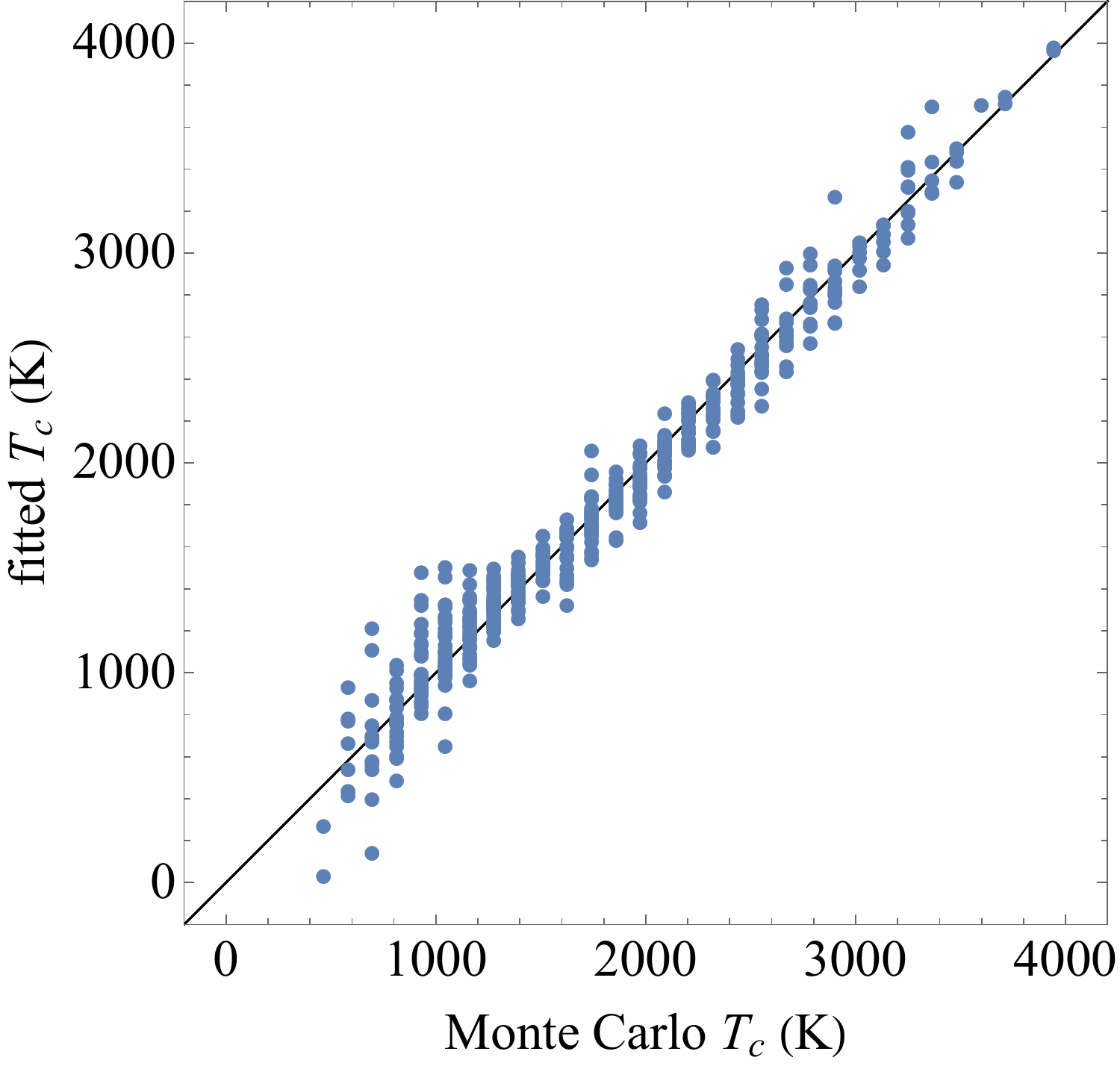}

\caption{A plot of the fitted $T_{c}$ against the $T_{c}$ from MC simulations
at 505 compositions. \label{fig:fitted-Tc-vs-MC-Tc}}
\end{figure}
\par\end{center}

Figure \ref{fig:fitted-Tc-vs-MC-Tc} compares the fitted $T_{c}$
with the $T_{c}$ from the MC simulations at 505 compositions. The
rms error of the fit is 1$28\K$, comparable to the temperature step
size in the MC simulations. The fit is noticeably better at higher
temperatures $T\gtrsim1400\K$, which include all the temperatures
of interest in the main text. Fitting the $T_{c}$ to a higher order
polynomial does not substantially improve the accuracy of the fit.
A more accuracy fit would require a more complex functional form containing
thermodynamic considerations.

\bibliographystyle{elsarticle-num-names}
\bibliography{library}

\end{document}